\documentclass[10pt]{IEEEtran}
\usepackage{graphicx}
\usepackage{tikz}
\usepackage{comment}
\usetikzlibrary{positioning,spy}
\usepackage{amsmath,amsfonts,amssymb,amsthm}
\usepackage{booktabs,tabularx}
\usepackage{cuted}
\usepackage[percent]{overpic}
\usepackage{lipsum}
\usepackage{tcolorbox}
\usepackage{multicol}
\usepackage{algorithm,algorithmic}
\usepackage{bm}
\usepackage{url}
\usepackage[inline]{enumitem}
\usepackage[caption=false,font=footnotesize]{subfig}
\usepackage[bottom]{footmisc}
\usepackage{fancyhdr}

\DeclareMathOperator*{\argmin}{arg\,min}

\def\bx{\ensuremath{{\bf x}}}
\def\bp{\ensuremath{{\bf p}}}
\def\bq{\ensuremath{{\bf q}}}
\def\bs{\ensuremath{{\bf s}}}
\def\bo{\ensuremath{{\bf o}}}
\def\bphi{\ensuremath{\bm{\Phi}}}
\def\balpha{\ensuremath{\bm{\alpha}}}
\def\bg{\ensuremath{{\bf g}}}
\def\by{\ensuremath{{\bf y}}}
\def\bz{\ensuremath{{\bf z}}}
\def\bw{\ensuremath{{\bf w}}}

\def\bW{\ensuremath{{\bf W}}}
\def\bK{\ensuremath{{\bf K}}}
\def\bM{\ensuremath{{\bf M}}}
\def\bk{\ensuremath{{\bf k}}}
\def\bn{\ensuremath{{\bf n}}}

\def\bb{\ensuremath{{\bf b}}}

\def\bP{\ensuremath{{\bf P}}}
\def\bQ{\ensuremath{{\bf Q}}}
\def\bR{\ensuremath{{\bf R}}}

\def\bR{\ensuremath{{\bf R}}}

\def\bU{\ensuremath{{\bf U}}}
\def\bV{\ensuremath{{\bf V}}}
\def\bD{\ensuremath{{\bf D}}}
\def\bI{\ensuremath{{\bf I}}}
\def\bL{\ensuremath{{\bf L}}}
\def\bS{\ensuremath{{\bf S}}}
\def\bX{\ensuremath{{\bf X}}}
\def\bH{\ensuremath{{\bf H}}}
\def\bN{\ensuremath{{\bf N}}}

\def\cP{\ensuremath{{\mathcal P}}}
\def\cT{\ensuremath{{\mathcal T}}}
\def\cF{\ensuremath{{\mathcal F}}}
\def\cM{\ensuremath{{\mathcal M}}}
\def\cN{\ensuremath{{\mathcal N}}}

\def\cS{\ensuremath{{\mathcal S}}}
\def\cH{\ensuremath{{\mathcal H}}}
\def\cV{\ensuremath{{\mathcal V}}}
\def\cE{\ensuremath{{\mathcal E}}}
\def\cU{\ensuremath{{\mathcal U}}}
\def\cP{\ensuremath{{\mathcal P}}}
\def\cL{\ensuremath{{\mathcal L}}}

\title{Algorithm Unrolling:  Interpretable, Efficient Deep Learning for Signal and Image Processing}
\author{Vishal~Monga,~\IEEEmembership{Senior~Member,~IEEE,} Yuelong Li,~\IEEEmembership{Member,~IEEE,}
	and Yonina C. Eldar,~\IEEEmembership{Fellow,~IEEE}

	\thanks{V. Monga is with Department of Electrical Engineering, The Pennsylvania State University, University Park,
		PA, 16802 USA, Email: vmonga@engr.psu.edu
	}
	\thanks{Y. Li is with Amazon Lab 126, San Jose, CA, 94089 USA, Email: liyuelongee@gmail.com}
	\thanks{Y. C. Eldar is with Weizmann  Institute  of  Science, Rehovot, Israel. Email: yonina.eldar@weizmann.ac.il.}

}

\begin{document}

\maketitle

\begin{abstract}
	Deep neural networks provide unprecedented performance gains in many real world problems in signal and image processing. 
	Despite these gains, future development and practical deployment of deep networks is hindered by their black-box nature, i.e., lack of interpretability, and by the need for very large training sets. An emerging technique called algorithm unrolling or unfolding offers promise in eliminating these issues by providing a concrete and systematic connection between iterative algorithms that are used widely in signal processing and deep neural networks. Unrolling methods were first proposed to develop fast neural network approximations for sparse coding. More recently, this direction has attracted enormous attention and is rapidly growing both in theoretic investigations and practical applications.  The growing popularity of unrolled deep networks is due in part to their  potential in developing efficient, high-performance and yet interpretable network architectures from reasonable size training sets. In this article, we review algorithm unrolling for signal and image processing. We extensively cover popular techniques for algorithm unrolling in various domains of signal and image processing including imaging, vision and recognition, and speech processing. By reviewing previous works, we reveal the connections between iterative algorithms and neural networks and present recent theoretical results. Finally, we provide a discussion on current limitations of unrolling and suggest possible future research directions.
\end{abstract}


\section{Introduction}\label{sec:intro}
The past decade has witnessed a deep learning revolution. Availability of
large-scale training datasets often facilitated by internet content, accessibility to
powerful computational resources thanks to breakthroughs in microelectronics,
and advances in neural network research such as development of effective
network architectures and efficient training algorithms have resulted in
unprecedented successes of deep learning in innumerable applications of computer
vision, pattern recognition and speech processing. For instance, deep learning has provided significant accuracy gains in image recognition, one of the core tasks in computer vision. Groundbreaking performance improvements have been demonstrated via
AlexNet~\cite{krizhevsky_imagenet_2012}, and lower classification errors than human-level
performance~\cite{he_delving_2015} was reported on the ImageNet
dataset~\cite{deng_imagenet_2009}.

In the realm of signal processing, learning based approaches provide an
interesting algorithmic alternative to traditional model based analytic
methods. In contrast to conventional iterative approaches where the models and
priors are typically designed by analyzing the physical processes and
handcrafting, deep learning approaches attempt to automatically discover model
information and incorporate them by optimizing network parameters that are
learned from real world training samples. Modern neural networks typically
adopt a hierarchical architecture composed of many layers and comprise a large
number of parameters (can be millions), and are thus capable of learning
complicated mappings which are difficult to design explicitly. When training
data is sufficient, this adaptivity enables deep networks to often overcome
model inadequacies, especially when the underlying physical scenario is hard to
characterize precisely. 

Another advantage of deep networks is that during inference, processing through
the network layers can be executed very fast. Many modern computational
platforms are highly optimized towards special operations such as convolutions,
so that inference via deep networks is usually quite computationally efficient.
In addition, the number of layers in a deep network is typically much smaller
than the number of iterations required in an iterative algorithm.  Therefore,
deep learning methods have emerged to offer desirable computational benefits over
state-of-the art in many areas of signal processing, imaging and vision. Their
popularity has reached new heights with widespread availability of the
supporting software infrastructure required for their implementation.

That said, a vast majority of deep learning techniques are
purely data-driven and the underlying structures are difficult to
interpret. Previous works largely apply general network architectures (some of
them will be covered in~\ref{ssec:conventional}) towards different problems,
and learn certain underlying mappings such as classification and regression
functions completely through end-to-end training. It is therefore hard to
discover what is learned inside the networks by examining the network
parameters, which are usually of high dimensionality, and what are the roles of
individual parameters. In other words, generic deep networks are usually difficult to interpret. In contrast, traditional iterative algorithms are usually highly interpretable because they are developed via modeling the physical processes underlying the problem and/or capturing prior domain knowledge. Interpretability is of course, an important
concern both in theory and practice. It is usually the key to conceptual
understanding and advancing the frontiers of knowledge. Moreover, in
areas such as medical applications and autonomous driving, it is crucial to
identify the limitations and potential failure cases of designed
systems, where interpretability plays a fundamental role.  Thus, lack of interpretability
can be a serious limitation of conventional deep learning methods
in contrast with model-based techniques with iterative algorithmic solutions which are used widely in signal processing.

An issue that frequently arises together with interpretability is
generalizability. It is well known that the practical success of deep learning
is sometimes overly dependent on the quantity and quality of training data
available. In scenarios where abundant high-quality training samples are
unavailable such as medical
imaging~\cite{Ronneberger_unet_2015,nabil_multires_2019} and 3D
reconstruction~\cite{nishida_procedural_2018}, the performance of deep networks
may degrade significantly, and sometimes may even underperform traditional
approaches. This phenomenon, formally called overfitting in machine learning
terminology, is largely due to the employment of generic neural networks that
are associated with a huge number of parameters. Without exploiting domain
knowledge explicitly beyond time and/or space invariance, such networks are highly under regularized and may
overfit severely even under heavy data augmentations.

To some extent, this
problem has recently been addressed via the design of domain enriched or prior
information guided deep networks~\cite{CellDetection, GuoSR, chen2018fsrnet, garnelo_neural_2018, sun2019functional}.
In such cases, the network architecture is designed to contain special layers
that are domain specific, such as the transform layer in~\cite{GuoSR}. In
other cases, prior structure of the expected output is exploited~\cite{GuoSR,
chen2018fsrnet} to develop training robustness. An excellent tutorial article
covering these issues is~\cite{lucas_using_2018}. Despite these achievements,
transferring domain knowledge to network parameters can be a nontrivial task
and effective priors may be hard to design. More importantly, the underlying
network architectures in these approaches largely remain consistent with
conventional neural networks. Therefore, the pursuit of
interpretable, generalizable and high performance deep architectures for signal
processing problems remains a highly important open challenge.

\begin{figure*}
	\includegraphics[width=\textwidth]{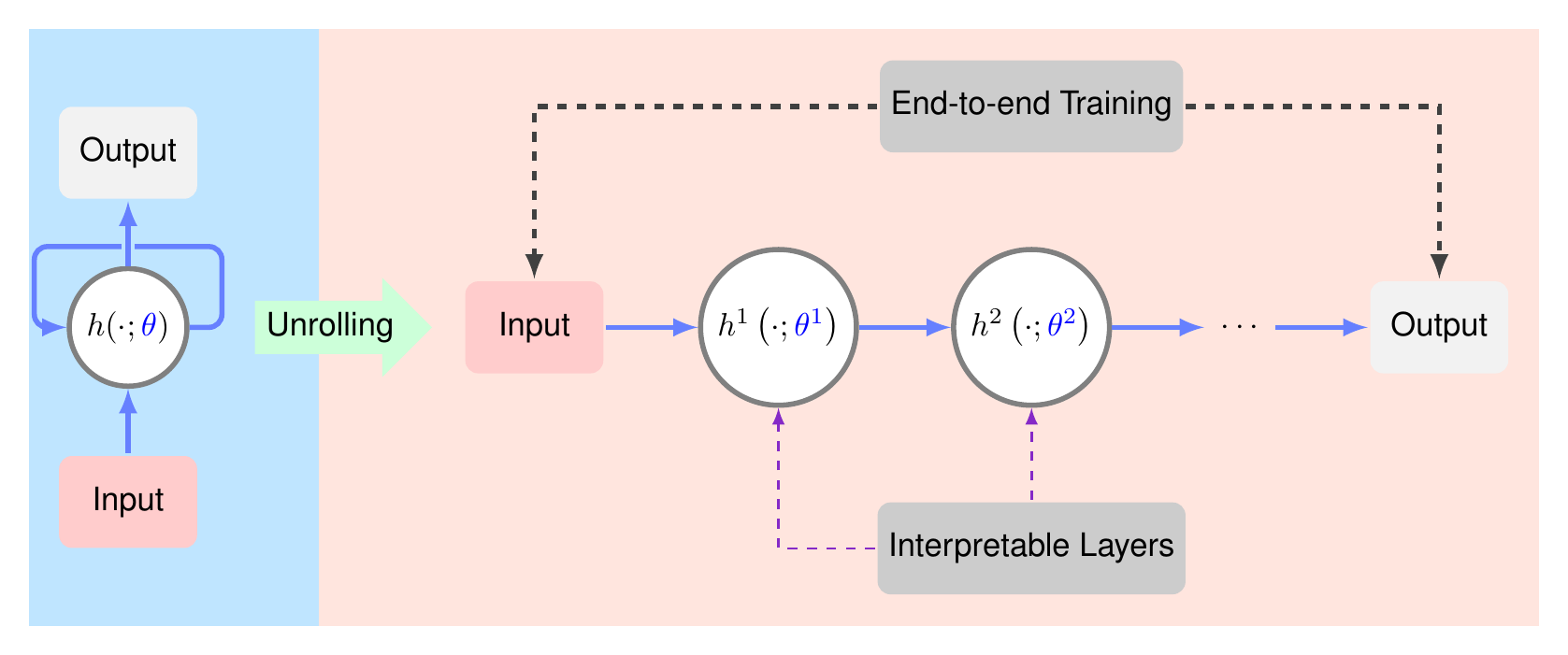}
	\caption{A high-level overview of algorithm unrolling: given an iterative algorithm (left), a corresponding deep network (right) can be generated by cascading its iterations $h$. The iteration step $h$ (left) is executed a number of times, resulting in the network layers $h^1, h^2, \dots$ (right). Each iteration $h$ depends on algorithm parameters $\theta$, which are transferred into network parameters $\theta^1,\theta^2,\dots$. Instead of determining these parameters through cross-validation or analytical derivations, we learn $\theta^1,\theta^2,\dots$ from training datasets through end-to-end training. In this way, the resulting network could achieve better performance than the original iterative algorithm. In addition, the network layers naturally inherit interpretability from the iteration procedure. The learnable parameters are colored in blue.}\label{fig:summary}
\end{figure*}

In the seminal work of Gregor and LeCun~\cite{gregor_learning_2010}, a
promising technique called algorithm unrolling (or unfolding) was developed that has helped
connect iterative algorithms such as those for sparse coding to neural network
architectures.  Following this work, the past few years have seen a surge of efforts
that unroll iterative algorithms for many significant problems in signal and
image processing: examples include (but are not limited to) compressive
sensing~\cite{yang_admm_csnet}, deconvolution~\cite{li_icassp19} and
variational techniques for image processing~\cite{chen_trainable_2017}.
Figure~\ref{fig:summary} provides a high-level illustration of this framework.
Specifically, each iteration of the algorithm step is represented as one layer of the
network. Concatenating these layers forms a deep neural network.  Passing
through the network is equivalent to executing the iterative algorithm a finite
number of times. In addition, the algorithm parameters (such as the model parameters and regularization coefficients) transfer to the network parameters. The network may be trained using back-propagation resulting in model parameters that are learned from real world training sets. In this way, the
trained network can be naturally interpreted as a parameter optimized algorithm,
effectively overcoming the lack of interpretability in most conventional neural
networks.  

Traditional iterative algorithms generally entail
significantly fewer parameters compared with popular neural networks. Therefore, the unrolled networks are highly parameter efficient and require less training data.
In addition, the unrolled networks naturally inherit prior structures and
domain knowledge rather than learn them from intensive training data.
Consequently, they tend to generalize better than generic networks, and can be computationally faster as long as each
algorithmic iteration (or the corresponding layer) is not overly expensive.

In this article, we review the foundations of
algorithm unrolling.  Our goal is to provide readers
with guidance on how to utilize unrolling to build efficient and interpretable neural networks in solving
signal and image processing problems. After providing a tutorial on how to unroll iterative algorithms into deep networks, we extensively review selected applications of algorithm unrolling in a wide variety of signal and image processing domains. We also review general theoretical studies
that shed light on convergence properties of these networks, although further analysis is an important problem for future research. In addition, we
clarify the connections between general classes of traditional iterative
algorithms and deep neural networks established through algorithm unrolling. We contrast algorithm unrolling with alternative approaches and
discuss their strengths and limitations. Finally, we discuss open challenges, and suggest future research directions.

\section{Generating Interpretable Networks through Algorithm Unrolling}\label{sec:unrolling}
We begin by describing algorithm unrolling. To motivate the unrolling approach, we commence with a brief review on conventional neural network architectures in Section~\ref{ssec:conventional}. We next discuss the first unrolling technique for sparse coding in Section~\ref{ssec:lista}. We elaborate on general forms of unrolling in Section~\ref{ssec:unroll_general}. 

\subsection{Conventional Neural Networks}\label{ssec:conventional}

\begin{figure}
	\centering
	\subfloat[\label{subfig:mlp}]{\includegraphics[width=\linewidth]{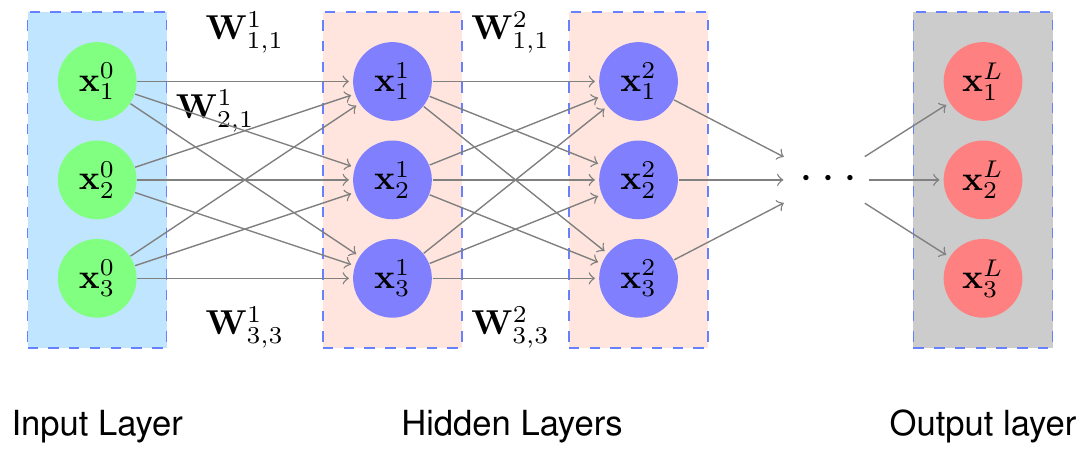}}\\
	\subfloat[\label{subfig:cnn}]{\includegraphics[width=\linewidth]{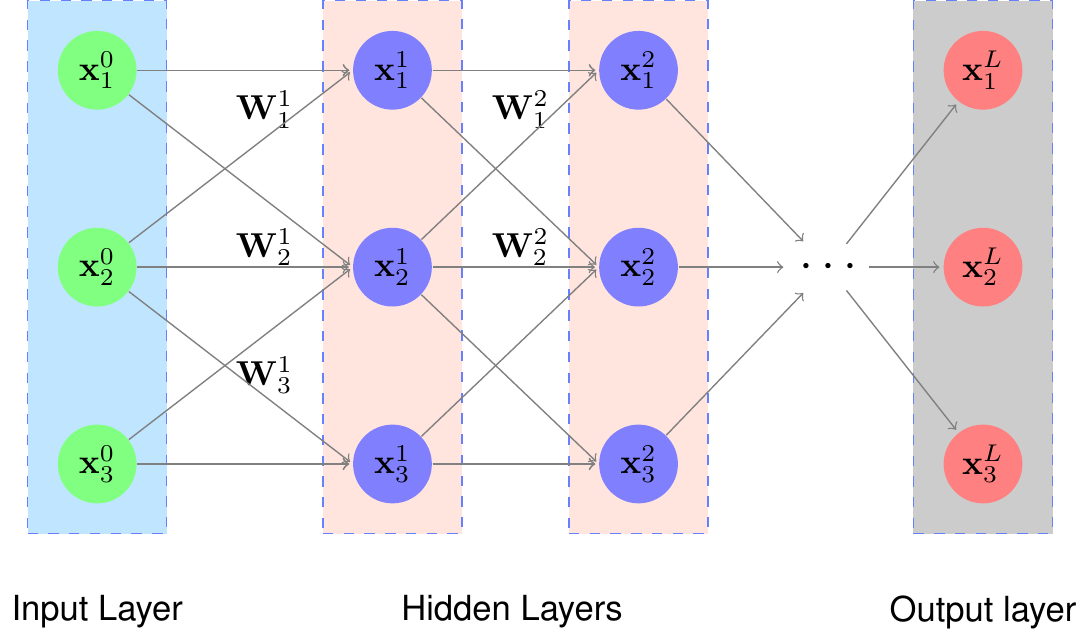}}\\
	\subfloat[\label{subfig:rnn}]{\includegraphics[width=\linewidth]{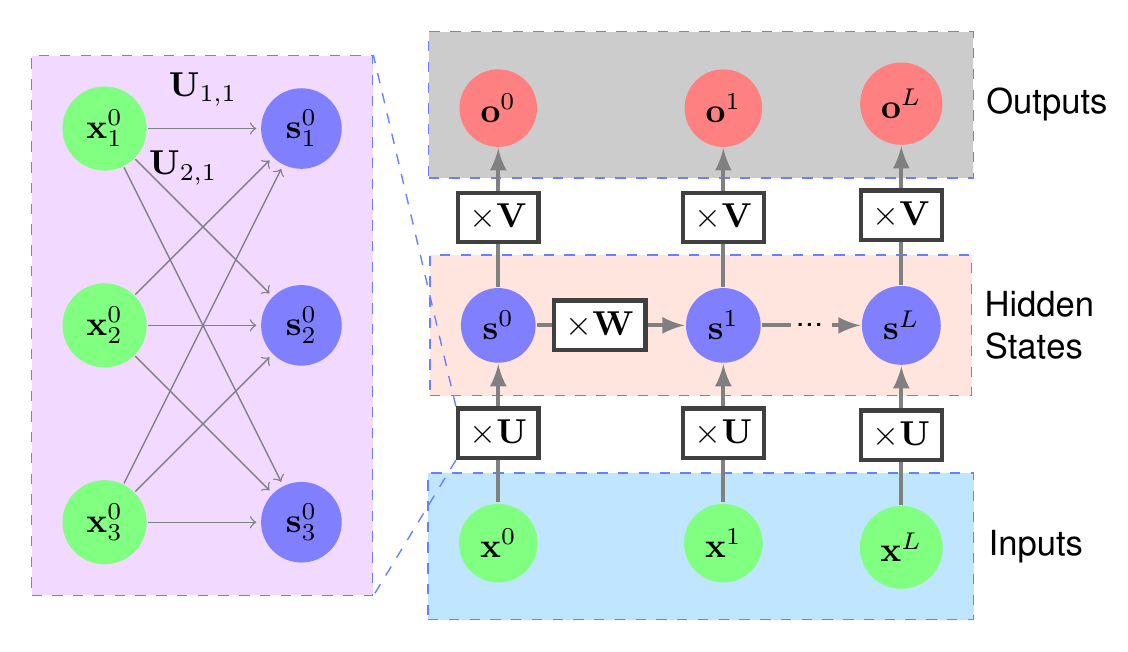}}\\
	\caption{Conventional neural network architectures that are popular in signal/image processing and computer vision applications: \protect\subref{subfig:mlp} Multi-Layer Perceptron (MLP) where all the neurons are fully connected; \protect\subref{subfig:cnn} Convolutional Neural Network (CNN) where the neurons are sparsely connected and the weights are shared between different neurons. Therefore, the weight matrices $\bW^l,l=1,2,\ldots,L$ effectively become convolution operators; \protect\subref{subfig:rnn} Recurrent Neural Network (RNN) where the inputs $\bx^l,l=1,2,\ldots,L$ are fed in sequentially and the parameters $\bU,\bV,\bW$ are shared across different time steps.}\label{fig:conventional}
\end{figure}

In early neural network research the Multi-Layer Perceptrons (MLP)
was a popular choice. This architecture can be motivated either biologically by
mimicking the human recognition system, or algorithmically by generalizing the
perceptron algorithm to multiple layers. A diagram illustration of MLP is
provided in Fig.~\ref{subfig:mlp}. The network is constructed through recursive linear
and nonlinear operations, which are called \emph{layers}. The units
those operations act upon are called \emph{neurons}, which is an analogy of the
neurons in the human nerve systems. The first and last layer are called \emph{input layer}
and \emph{output layer}, respectively.

A salient property of this network is that each neuron is fully connected to
every neuron in the previous layer, except for the input layer. The layers are
thus commonly called Fully-Connected layers. Analytically, in the $l$-th
layer the relationship between the neurons $\bx_j^l$ and $\bx_i^{l+1}$ is
expressed as
\begin{equation}
	\bx_i^{l+1}=\sigma\left(\sum_j\bW_{ij}^{l+1}\bx_j^l+\bb_i^{l+1}\right)\label{eq:cnn_layer}
\end{equation}    
where
$\bW^{l+1}$ and $\bb^{l+1}$ are the \emph{weights} and
\emph{biases}, respectively, and $\sigma$ is a nonlinear \emph{activation function}. We omit drawing activation
functions and biases in Fig.~\ref{subfig:mlp} for brevity. Popular choices of
activation functions include the logistic function and the hyperbolic tangent
function.  In recent years, they have been superseded by Rectified Linear
Units (ReLU)~\cite{glorot2011deep} defined by 
\begin{equation*}
    \mathrm{ReLU}(x)=\max\{x,0\}.
\end{equation*}
The $\bW$'s and $\bb$'s are generally trainable parameters that are learned from
datasets through training, during which
back-propagation~\cite{lecun_efficient_2012} is often employed for gradient
computation.

Nowadays, MLPs are rarely seen in practical imaging and vision applications.
The fully-connected nature of MLPs contributes to a rapid increase in their parameters,
making them difficult to train. To address this limitation, Fukushima {\it et al.}~\cite{fukushima_neocognitron_1980} designed a neural network by mimicking the visual nervous system~\cite{hubel_receptive_1962}.
The neuron connections are restricted
to local neighbors only and weights are shared across different spatial locations. The linear operations then become
convolutions (or correlations in a strict sense) and thus the networks
employing such localizing structures are generally called Convolutional Neural
Networks (CNN). A visual illustration of a CNN can be seen in Fig.~\ref{subfig:cnn}. With
significantly reduced parameter dimensionality, training deeper networks
becomes feasible. While CNNs were first applied to digit recognition,
their translation invariance is a desirable property in many
computer vision tasks. CNNs thus have become an extremely popular and indispensable
architecture in imaging and vision, and outperform traditional approaches
by a large margin in many domains. Today they continue to exhibit the best performance in many applications.

In domains such as speech recognition and video processing, where data
is obtained sequentially, Recurrent Neural Networks
(RNN)~\cite{rumelhart_learning_1986} are a popular choice. RNNs explicitly
model the data dependence in different time steps in the sequence, and scale
well to sequences with varying lengths. A visual depiction of RNNs is provided in
Fig.~\ref{subfig:rnn}. Given the previous hidden state $\bs^{l-1}$ and input variable $\bx^l$, the next hidden state $\bs^l$ is computed as
\begin{equation*}
	\bs^l=\sigma_1\left(\bW\bs^{l-1}+\bU\bx^l+\bb\right),
\end{equation*}
while the output variable $\bo^l$ is generated by
\begin{equation*}
	\bo^l=\sigma_2\left(\bV\bs^l+\bb\right).
\end{equation*}
Here $\bU,\bV,\bW,\bb$ are trainable network parameters and $\sigma_1,\sigma_2$ are activation functions. We again omit the activation functions and biases in Fig.~\ref{subfig:rnn}. In contrast to MLPs and CNNs where the layer operations
are applied recursively in a hierarchical representation fashion, RNNs apply
the recursive operations as the time step evolves. A distinctive property of
RNNs is that the parameters $\bU, \bV, \bW$ are shared across all the time steps, rather than varying from layer to layer. Training RNNs
can thus be difficult as the gradients of the parameters may either explode or
vanish.

\subsection{Unrolling Sparse Coding Algorithms into Deep Networks}\label{ssec:lista}

The earliest work in algorithm unrolling dates back to Gregor {\it et al.}'s
paper on improving the computational efficiency of sparse coding algorithms
through end-to-end training~\cite{gregor_learning_2010}. In particular, they
discussed how to improve the efficiency of the Iterative Shrinkage and
Thresholding Algorithm (ISTA), one of the most popular approaches in sparse
coding. The crux of this work is summarized in Fig.~\ref{fig:lista} and detailed in the box on Learned ISTA.\@ Each
iteration of ISTA comprises one linear operation followed by a non-linear
soft-thresholding operation, which mimics the ReLU activation function.
A diagram representation of one iteration step reveals its resemblance
to a single network layer. Thus, one can form a deep network by mapping each iteration to a network layer and
stacking the layers together which is equivalent to executing an ISTA
iteration multiple times. Because
the same parameters are shared across different layers, the resulting network
resembles an RNN in terms of architecture. In recent studies~\cite{xin2016maximal,liu2018alista,chen_theoretical_2018}, different parameters are employed in different layers, as we discuss in Section~\ref{ssec:theory}.

After unrolling ISTA into a network, the network is trained using
training samples through back-propagation. The learned network is dubbed
Learned ISTA (LISTA). It turns out that significant computational benefits can
be obtained by learning from real data. For instance, Gregor {\it et
al.}~\cite{gregor_learning_2010} experimentally verified that the learned network
reaches a specific performance level around $20$ times faster than
accelerated ISTA.\@ Consequently, the sparse coding problem can be solved
efficiently by passing through a compact LISTA network.
From a theoretical perspective, recent
studies~\cite{liu2018alista,chen_theoretical_2018} have characterized the
linear convergence rate of LISTA and further verified its computational
advantages in a rigorous and quantitative manner. A more detailed exposition and
discussion on related theoretical studies will be provided in
Section~\ref{ssec:theory}.

In addition to ISTA, Gregor {\it et al.} discussed unrolling and
optimizing another sparse coding method, the Coordinate Descent (CoD)
algorithm~\cite{li_coordinate_2009}. The technique and implications of unrolled
CoD are largely similar to LISTA.\@

\subsection{Algorithm Unrolling in General}\label{ssec:unroll_general}

\begin{figure*}
	\centering
	\includegraphics[width=\textwidth]{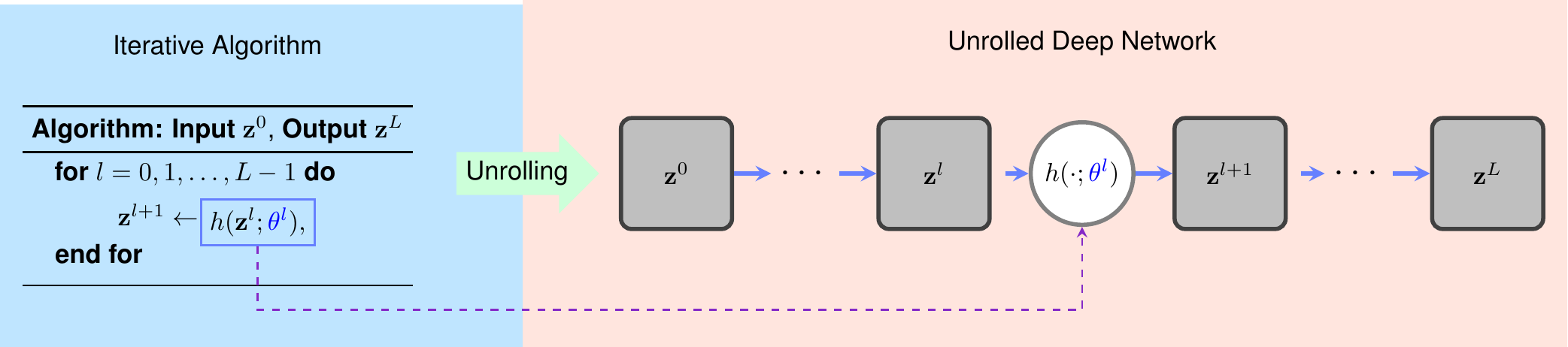}
	\caption{Illustration of the general idea of algorithm unrolling: starting with an abstract iterative algorithm, we map one iteration (described as the function $h$ parametrized by $\theta^l,l=0,\dots,L-1$) into a single network layer, and stack a finite number of layers together to form a deep network. Feeding the data forward through an $L$-layer network is equivalent to executing the iteration $L$ times (finite truncation). The parameters $\theta^l,l=0,1,\dots,L-1$ are learned from real datasets by training the network end-to-end to optimize the performance. They can either be shared across different layers or varying from layer to layer. The trainable parameters are colored in blue.}\label{fig:unroll_general}
\end{figure*}

Although Gregor {\it et al.}'s work~\cite{gregor_learning_2010} focused on improving computational efficiency of sparse coding, the same techniques can be applied to general iterative algorithms.
An illustration is given in Fig.~\ref{fig:unroll_general}. In
general, the algorithm repetitively performs certain analytic operations, which
we represent abstractly as the $h$ function. Similar to LISTA, we unroll
the algorithm into a deep network by mapping each iteration into a single
network layer and stacking a finite number of layers together. Each iteration
step of the algorithm contains parameters, such as the model
parameters or the regularization coefficients, which we denote by vectors $\theta^l,l=0,\dots,L-1$. Through unrolling, the
parameters $\theta^l$ correspond to those of the
deep network, and can be optimized towards real-world scenarios by training
the network end-to-end using real datasets.

While the motivation of LISTA was computational savings, proper
use of algorithm unrolling can also lead to dramatically improved performance
in practical applications. For instance, we can employ back-propagation to
obtain coefficients of filters~\cite{li_icassp19} and
dictionaries~\cite{wang_deep_2015} that are hard to design either analytically
or even by handcrafting. In addition, custom modifications may be employed in the
unrolled network~\cite{yang_admm_csnet}. As a particular example, in LISTA (see the box Learned ISTA), the matrices $\bW_t$, $\bW_e$ may be learned in each iteration, so that they are no longer held fixed throughout the network. Furthermore, their values may vary across different layers rather than being shared. By allowing a slight departure from the
original iterative algorithms~\cite{gregor_learning_2010,jin_deep_2017} and
extending the representation capacity, the performance of the unrolled networks
may be boosted significantly.

Compared with conventional generic neural networks, unrolled networks
generally contain significantly fewer parameters, as they encode domain
knowledge through unrolling. In addition, their structures are more specifically tailored
towards target applications. These benefits not only ensure higher
efficiency, but also provide better generalizability especially under limited
training schemes~\cite{yang_admm_csnet}. More concrete examples will be
presented and discussed in Section~\ref{sec:applications}.

\section{Unrolling in Signal and Image Processing Problems}\label{sec:applications}

Algorithm unrolling has been applied to diverse application areas in the past few years. Table~\ref{tab:summary_app} summarizes representative methods and their topics of focus in different domains. Evidently, research in algorithm unrolling is growing and influencing a variety of high impact real-world problems and research areas. As discussed in Section~\ref{sec:unrolling}, an essential element of each unrolling approach is the underlying iterative algorithm it starts from, which we also specify in Table~\ref{tab:summary_app}.

In this section we discuss a variety of practical applications of algorithm unrolling. Specifically, we cover applications in computational imaging, medical imaging, vision and recognition, and other signal processing topics from Section~\ref{subsec:app_comput_imag} to Section~\ref{subsec:app_speech} sequentially. We then discuss the enhanced efficiency brought about by algorithm unrolling. 

\begin{figure*}
	\centering
	\includegraphics[width=\textwidth]{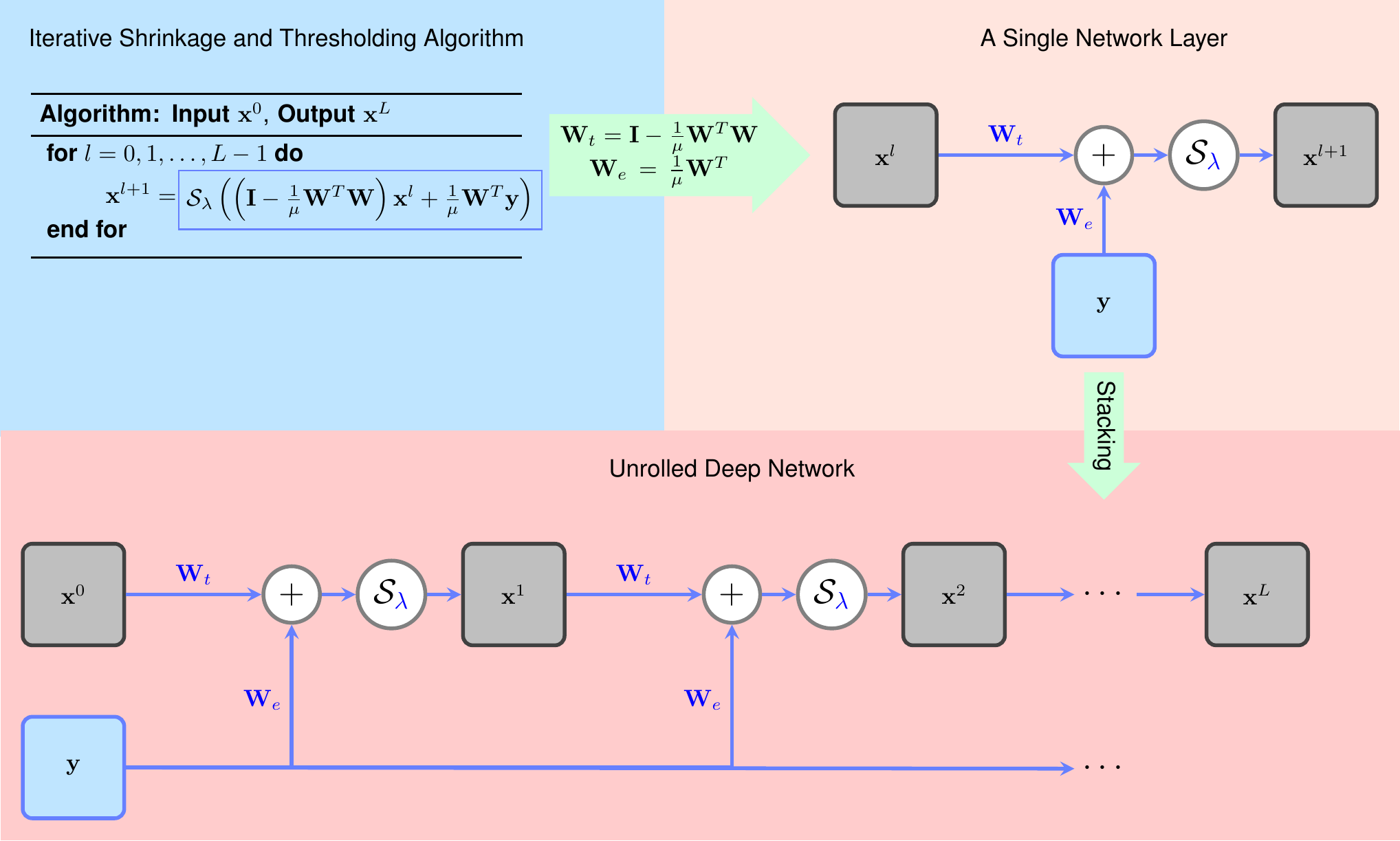}
	\caption{Illustration of LISTA:\@ one iteration of ISTA executes a linear and then a non-linear operation and thus can be recast into a network layer; by stacking the layers together a deep network is formed. The network is subsequently trained using paired inputs and outputs by back-propagation to optimize the parameters $\bW_e,\bW_t$ and $\lambda$. $\mu$ is a constant parameter that controls the step size of each iteration. The trained network, dubbed LISTA, is computationally more efficient compared with the original ISTA.\@ The trainable parameters in the network are colored in blue. For details, see the box on LISTA.\@ In practice, $\bW_e,\bW_t,\lambda$ may vary in each layer.}\label{fig:lista}
\end{figure*}

\begin{strip}
\begin{tcolorbox}[title={Learned ISTA},parbox=false]
	\begin{multicols}{2}
	The pursuit of parsimonious representation of signals has been a problem of enduring interest in signal processing. One of the most common quantitative manifestations of this is the well-known sparse coding problem~\cite{eldar_2012_compressed}. Given an input vector $\by\in\mathbb{R}^n$ and an over-complete dictionary $\bW\in\mathbb{R}^{n\times m}$ with $m>n$, sparse coding refers to the pursuit of a sparse representation of $\by$ using $\bW$. In other words, we seek a sparse code $\bx\in\mathbb{R}^m$ such that $\by\approx\bW\bx$ while encouraging as many coefficients in $\bx$ to be zero (or small in magnitude) as possible. A well-known approach to determine $\bx$ is to solve the following convex optimization problem:
	\begin{align}
		\min_{\bx\in\mathbb{R}^m}\frac{1}{2}\|\by-\bW\bx\|_2^2+\lambda\|\bx\|_1,\label{eqn:l1_min}
	\end{align}
	where $\lambda>0$ is a regularization parameter that controls the sparseness of the solution.
	
	A popular method for solving~\eqref{eqn:l1_min} is the family of Iterative Shrinkage and Thresholding Algorithms (ISTA)~\cite{beck2009fast}. In its simplest form, ISTA performs the following iterations:
	\begin{align}
		\bx^{l+1}=\cS_{\lambda}\left\{\left(\bI-\frac{1}{\mu}{\bW}^T\bW\right)\bx^l+\frac{1}{\mu}\bW^T\by\right\},\, l=0,1,\dots\label{eqn:ista}
	\end{align}
	where $\mathbf{I}\in\mathbb{R}^{m\times m}$ is the identity matrix, $\mu$ is a positive parameter that controls the iteration step size, $\cS_\lambda(\cdot)$ is the soft-thresholding operator defined elementwise as
	\begin{equation}
		\cS_\lambda(x)=\mathrm{sign}(x)\cdot\max\left\{|x|-\lambda, 0\right\},\label{eq:soft_thresh}
    \end{equation}
    Basically, ISTA is equivalent to a gradient step of $\|\by-\bW\bx\|_2^2$, followed by a projection onto the $\ell^1$ ball.

	As depicted in Fig.~\ref{fig:lista}, the iteration~\eqref{eqn:ista} can
	be recast into a single network layer. This layer comprises a
	series of analytic operations (matrix-vector multiplication, summation,
	soft-thresholding), which is of the same nature as a neural network.
	Executing ISTA $L$ times can be interpreted as cascading $L$ such layers
	together, which essentially forms an $L$-layer deep network. In the unrolled network an implicit substitution of parameters has been made: $\bW_t=\bI-\frac{1}{\mu}\bW^T\bW$ and $\bW_e=\frac{1}{\mu}\bW^T$.
	While these substitutions generalize the original parametrization and
	expand the representation power of the unrolled network, recent theoretical
	studies~\cite{chen_theoretical_2018} suggest that they may
	be inconsequential in an asymptotic sense as the optimal network parameters
	admit a weight coupling scheme asymptotically.

	The unrolled network is trained using real datasets to optimize the parameters $\bW_t$, $\bW_e$ and $\lambda$. The learned version, LISTA, may achieve higher efficiency compared to ISTA.\@ It is also useful when $\bW$ is not known exactly.  Training is performed using a sequence of vectors $\by^1,\by^2,\dots,\by^N\in\mathbf{R}^n$ and their corresponding groundtruth sparse codes ${\bx^\ast}^1,{\bx^\ast}^2,\dots,{\bx^\ast}^N$.
	
	By feeding each $\by^n,n=1,\dots,N$ into the network, we retrieve its output $\widehat{\bx}^n\left(\by^n;\bW_t,\bW_e,\lambda\right)$ as the predicted sparse code for $\by^n$. Comparing it with the ground truth sparse code ${\bx^\ast}^n$, the network training loss function is formed as:

	\begin{align}
		\ell(\bW_t,\bW_e,\lambda)=\frac{1}{N}\sum_{n=1}^N\left\|\widehat{\bx}^n\left(\by^n;\bW_t,\bW_e,\lambda\right)-{\bx^\ast}^n\right\|_2^2,\label{eqn:lista_loss}
	\end{align}
	and the network is trained through loss minimization, using popular gradient-based learning techniques such as stochastic gradient descent~\cite{lecun_efficient_2012}, to learn $\bW_t$, $\bW_t$, $\lambda$. It has been shown empirically that, the number of layers $L$ in (trained) LISTA can be an order of magnitude smaller than the number of iterations required for ISTA~\cite{gregor_learning_2010} to achieve convergence corresponding to a new observed input.
	\end{multicols}
\end{tcolorbox}
\end{strip}

\begin{table*}
	\centering
	\caption{Summary of Recent Methods Employing Algorithm Unrolling in Practical Signal Processing and Imaging Applications.}\label{tab:summary_app}
	\begin{tabularx}{\textwidth}{c c c X X}
		\toprule
		Reference & Year & Application domain & Topics & Underlying Iterative Algorithms\\
		\midrule
		Hershey {\it et al.}~\cite{hershey_2014_deep} & 2014 & Speech Processing & Signal channel source separation & Non-negative matrix factorization\\
		\midrule
		Wang {\it et al.}~\cite{wang_deep_2015} & 2015 & Computational imaging & Image super-resolution & Coupled sparse coding with iterative shrinkage and thresholding\\
		\midrule
		Zheng {\it et al.}~\cite{zheng_conditional_2015} & 2015 & Vision and Recognition & Semantic image segmentation & Conditional random field with mean-field iteration\\
		\midrule
		Schuler {\it et al.}~\cite{schuler_learning_2016} & 2016 & Computational imaging & Blind image deblurring & Alternating minimization\\
		\midrule
		Chen {\it et al.}~\cite{chen_trainable_2017} & 2017 & Computational imaging & Image denoising, JPEG deblocking & Nonlinear diffusion\\
		\midrule
		Jin {\it et al.}~\cite{jin_deep_2017} & 2017 & Medical Imaging & Sparse-view X-ray computed tomography & Iterative shrinkage and thresholding\\
		\midrule
		Liu {\it et al.}~\cite{liu_deep_2018} & 2018 & Vision and Recognition & Semantic image segmentation & Conditional random field with mean-field iteration\\
		\midrule
		Solomon {\it et al.}~\cite{solomon_deep_2018} & 2018 & Medical imaging & Clutter suppression & Generalized ISTA for robust principal component analysis\\
		\midrule
		Ding {\it et al.}~\cite{ding_domain_2018} & 2018 & Computational imaging & Rain removal & Alternating direction method of multipliers\\
		\midrule
		Wang {\it et al.}~\cite{wang_2018_end} & 2018 & Speech processing & Source separation & Multiple input spectrogram inversion\\
		\midrule
		Adler {\it et al.}~\cite{adler2018learned} & 2018 & Medical Imaging & Computational tomography & Proximal dual hybrid gradient\\
		\midrule
		Wu {\it et al.}~\cite{wu_end--end_2018} & 2018 & Medical Imaging & Lung nodule detection & Proximal dual hybrid gradient\\
		\midrule
		Yang {\it et al.}~\cite{yang_admm_csnet} & 2019 & Medical imaging & Medical resonance imaging, compressive imaging & Alternating direction method of multipliers\\
		\midrule
		Hosseini {\it et al.}~\cite{hosseini_dense_2019} & 2019 & Medical imaging & Medical resonance imaging & Proximal gradient descent\\
		\midrule
		Li {\it et al.}~\cite{li_2019_deep} & 2019 & Computational imaging & Blind image deblurring & Half quadratic splitting\\
		\midrule
		Zhang {\it et al.}~\cite{zhang_real-time_2019} & 2019 & Smart power grids & Power system state estimation and forecasting & Double-loop prox-linear iterations\\
		\midrule
		Zhang {\it et al.}~\cite{zhang_dynamically_2018} & 2019 & Computational imaging & Blind image denoising, JPEG deblocking & Moving endpoint control problem\\
		\midrule
		Lohit {\it et al.}~\cite{lohit_unrolled_2019} & 2019 & Remote sensing & Multi-spectral image fusion & Projected gradient descent\\
		\midrule
		Yoffe {\it et al.}~\cite{dardikman-yoffe_learned_2020} & 2020 & Medical Imaging & Super resolution microscopy & Sparsity-based super-resolution microscopy from correlation information~\cite{solomon_sparsity-based_2018}\\
		\bottomrule
	\end{tabularx}
\end{table*}

\begin{figure*}
	\centering
	\includegraphics[width=\linewidth]{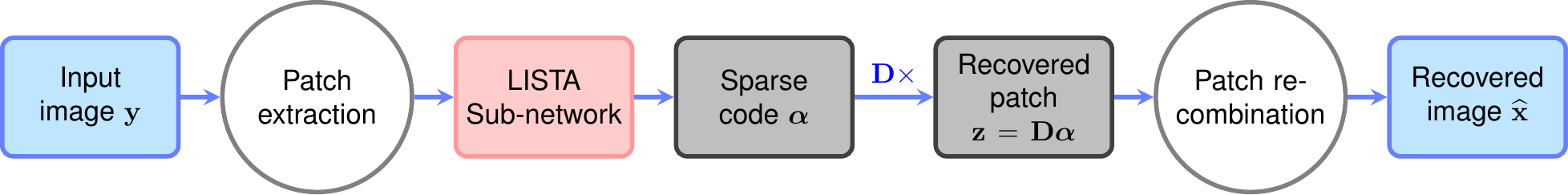}
	\caption{Illustration of the SCN~\cite{wang_deep_2015} architecture: the patches extracted from input low-resolution image $\by$ are fed into a LISTA sub-network to estimate the associated sparse codes $\bm{\alpha}$, and then high-resolution patches are reconstructed through a linear layer. The predicted high-resolution image $\widehat{\bx}$ is formed by putting these patches into their corresponding spatial locations. The whole network is trained by forming low-resolution and high-resolution image pairs, using standard stochastic gradient descent algorithm. The high resolution dictionary $\bD$ (colored in blue) and the LISTA parameters are trainable from real datasets.}\label{fig:scn}
\end{figure*}

\begin{figure*}
	\centering
	\begin{tikzpicture}[spy using outlines={rectangle,green,magnification=8,size=1.5cm}]
		\node[inner sep=0em] (baby_gt){\pgfimage[height=3.8cm]{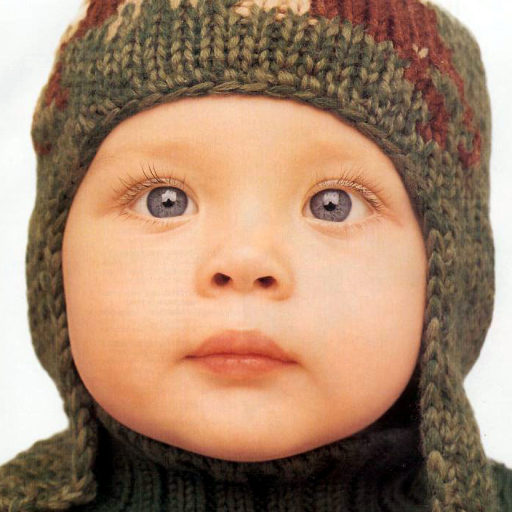}};
		\node[inner sep=0em,right=0.1cm of baby_gt] (bird_gt){\pgfimage[height=3.8cm]{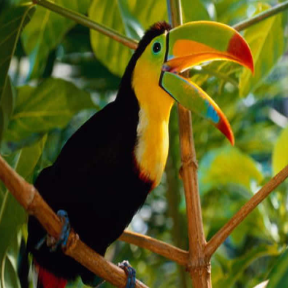}};
		\spy on (3.9,1.2) in node [left] at (5.6,-1);
		\node[inner sep=0em,right=0.1cm of bird_gt] (butterfly_gt){\pgfimage[height=3.8cm]{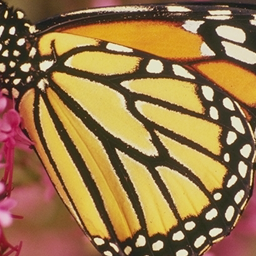}};
		\spy on (7,0.45) in node [left] at (9.5,-1);
		\node[inner sep=0em,right=0.1cm of butterfly_gt] (head_gt){\pgfimage[height=3.8cm]{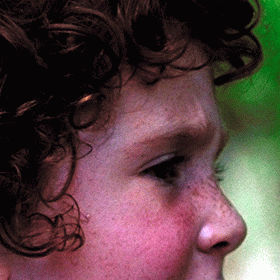}};
		\node[inner sep=0em,right=0.1cm of head_gt] (woman_gt){\pgfimage[height=3.8cm]{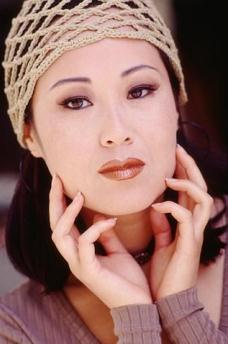}};
		\spy on (14.5,1) in node [left] at (16,-1);
		\node[inner sep=0em,below=0.2cm of baby_gt] (baby_aplus){\pgfimage[height=3.8cm]{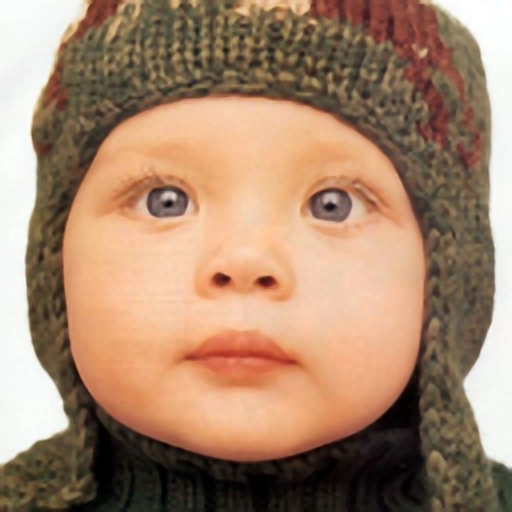}};
		\node[inner sep=0em,right=0.1cm of baby_aplus] (bird_aplus){\pgfimage[height=3.8cm]{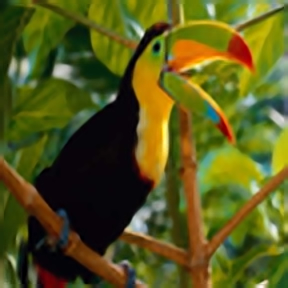}};
		\spy on (3.9,-2.8) in node [left] at (5.6,-5);
		\node[inner sep=0em,right=0.1cm of bird_aplus] (butterfly_aplus){\pgfimage[height=3.8cm]{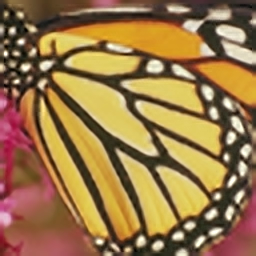}};
		\spy on (7,-3.6) in node [left] at (9.5,-5);
		\node[inner sep=0em,right=0.1cm of butterfly_aplus] (head_aplus){\pgfimage[height=3.8cm]{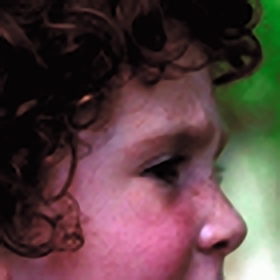}};
		\node[inner sep=0em,right=0.1cm of head_aplus] (woman_aplus){\pgfimage[height=3.8cm]{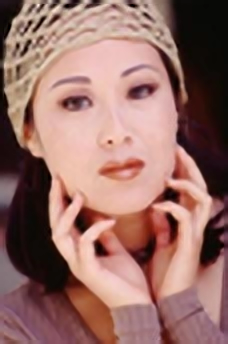}};
		\spy on (14.5,-3) in node [left] at (16,-5);
		\node[inner sep=0em,below=0.2cm of baby_aplus] (baby_cnn){\pgfimage[height=3.8cm]{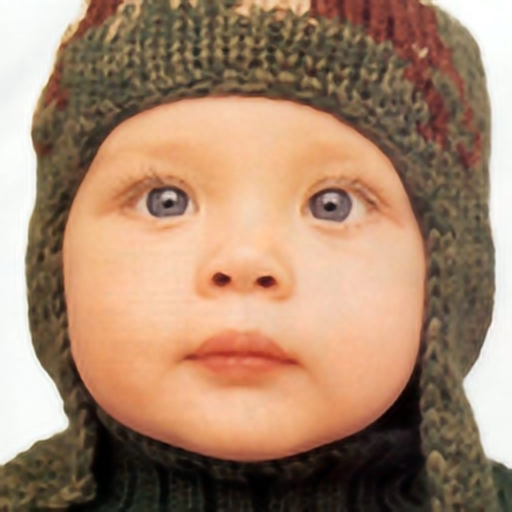}};
		\node[inner sep=0em,right=0.1cm of baby_cnn] (bird_cnn){\pgfimage[height=3.8cm]{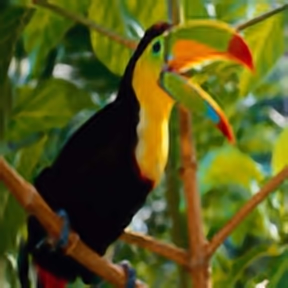}};
		\spy on (3.9,-6.8) in node [left] at (5.6,-9);
		\node[inner sep=0em,right=0.1cm of bird_cnn] (butterfly_cnn){\pgfimage[height=3.8cm]{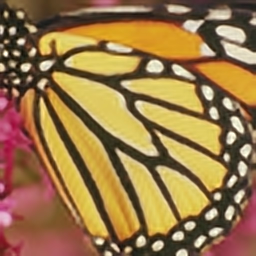}};
		\node[inner sep=0em,right=0.1cm of butterfly_cnn] (head_cnn){\pgfimage[height=3.8cm]{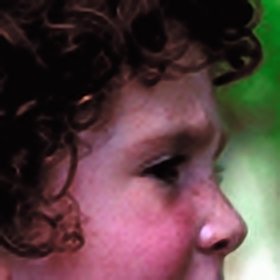}};
		\spy on (7,-7.6) in node [left] at (9.5,-9);
		\node[inner sep=0em,right=0.1cm of head_cnn] (woman_cnn){\pgfimage[height=3.8cm]{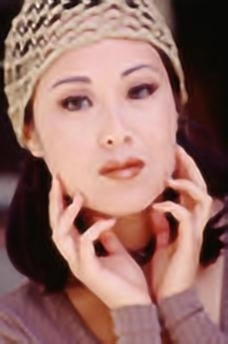}};
		\spy on (14.5,-7.03) in node [left] at (16,-9);
		\node[inner sep=0em,below=0.2cm of baby_cnn] (baby_scn){\pgfimage[height=3.8cm]{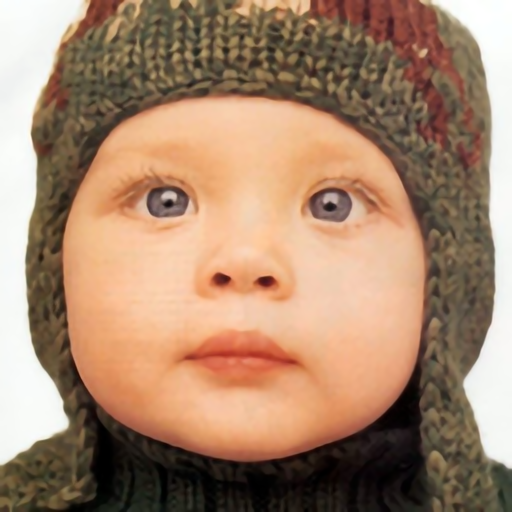}};
		\node[inner sep=0em,right=0.1cm of baby_scn] (bird_scn){\pgfimage[height=3.8cm]{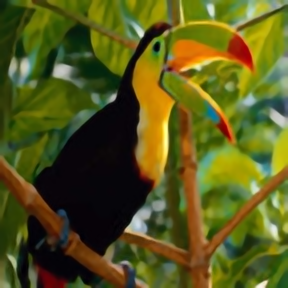}};
		\spy on (3.9,-10.8) in node [left] at (5.6,-13);
		\node[inner sep=0em,right=0.1cm of bird_scn] (butterfly_scn){\pgfimage[height=3.8cm]{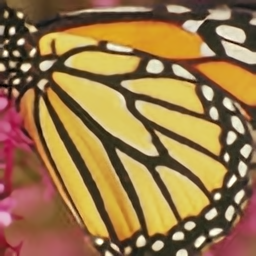}};
		\spy on (7,-11.6) in node [left] at (9.5,-13);
		\node[inner sep=0em,right=0.1cm of butterfly_scn] (head_scn){\pgfimage[height=3.8cm]{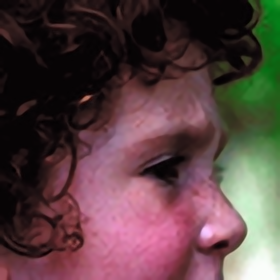}};
		\node[inner sep=0em,right=0.1cm of head_scn] (woman_scn){\pgfimage[height=3.8cm]{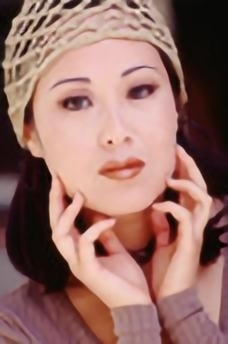}};
		\spy on (14.5,-11.03) in node [left] at (16,-13);
	\end{tikzpicture}
	\caption{Sample experimental results from~\cite{wang_deep_2015} for visual comparison on single image super-resolution. Groundtruth images are in the top row. The second to the bottom rows include results from~\cite{timofte2014a+},~\cite{dong_super_2016} and~\cite{wang_deep_2015}, respectively. These include a state-of-the art iterative algorithm as well as a deep learning technique. Note that the magnified portions show that SCN better recovers sharp edges and spatial details.}\label{fig:scn_set5}
\end{figure*}

\begin{figure*}
	\includegraphics[width=\textwidth]{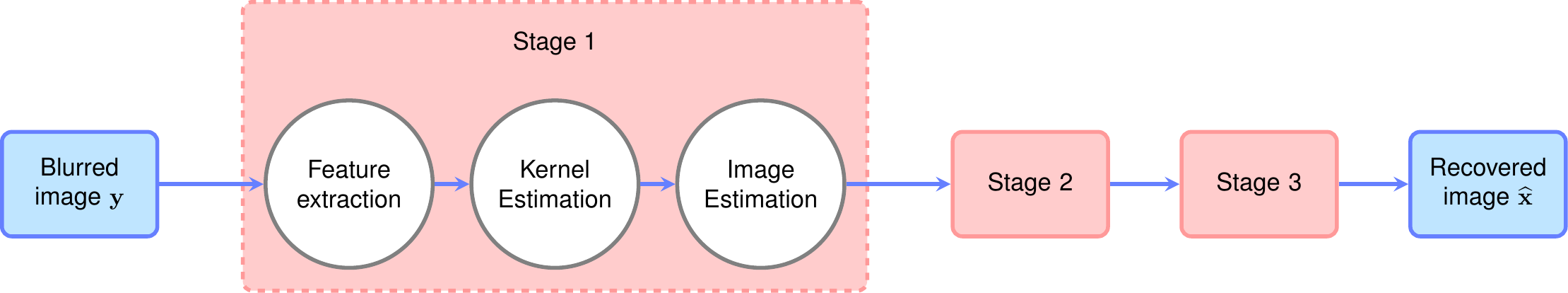}
	\caption{Illustration of the network architecture in~\cite{schuler_learning_2016}: the network is formed by concatenating multiple stages of essential blind image deblurring modules. Stage 2 and 3 repeats the same operations as stage 1, with different trainable parameters. From a conceptual standpoint, each stage imitates one iteration of a typical blind image deblurring algorithm. The training data can be formed by synthetically blurring sharp images to obtain their blurred versions.}\label{fig:scheuler}
\end{figure*}

\subsection{Applications in Computational Imaging}\label{subsec:app_comput_imag}
\begin{figure*}
	\centering
	\includegraphics[width=\textwidth]{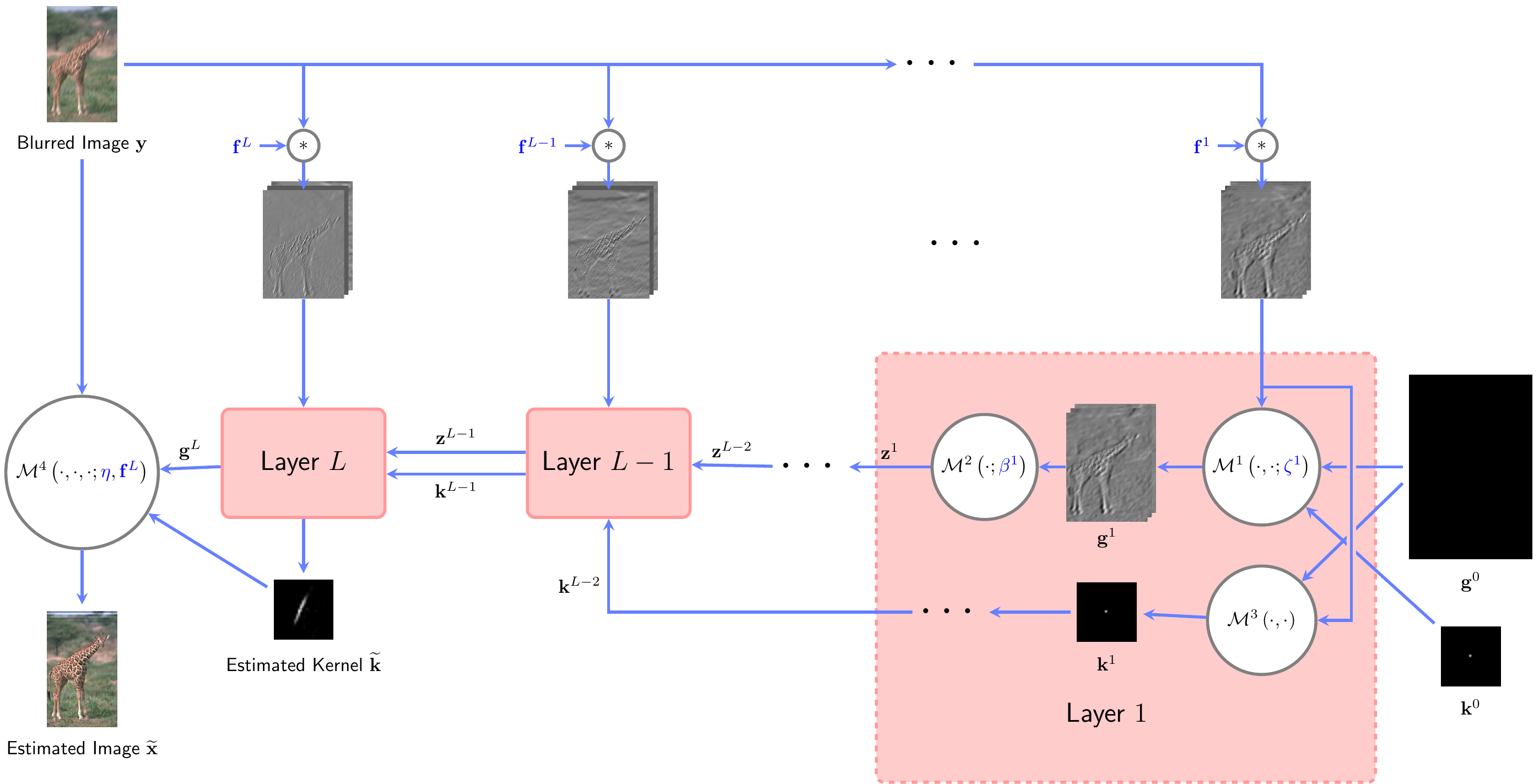}
	\caption{Diagram illustration of DUBLID~\cite{li_icassp19}. The analytical operations $\cM^1,\cM^2,\cM^3,\cM^4$ correspond to casting the analytic expressions in~\eqref{eq:analytic_expressions} and~\eqref{eq:image_retrieval} into the network. Trainable parameters are colored in blue. In particular, the parameters $\mathbf{f}^l, l=1,\dots,L$ denote trainable filter coefficients in the $l$-th layer.}\label{fig:dublid}
\end{figure*}

Computational imaging is a broad area covering a wide range of interesting
topics, such as computational photography, hyperspectral imaging, and compressive
imaging, to name a few. The key to success in many computational imaging areas
frequently hinges on solving an inverse problem. Model-based
inversion has long been a popular approach. Examples of model-based methods
include parsimonious representations such as sparse coding and low-rank matrix pursuit,
variational methods and conditional random fields. The employment of
model-based techniques gives rise to many iterative approaches, as closed-form
solutions are rarely available. The fertile ground of these iterative
algorithms in turn provides a solid foundation and offers many opportunities
for algorithm unrolling.

Single image super-resolution is an important topic in computational imaging
that focuses on improving the spatial resolution of a single degraded image. In
addition to offering images of improved visual quality, super-resolution also
aids diagnosis in medical applications and promises to improve the performance
of recognition systems.  Compared with naive bicubic interpolation, there exists large room for performance improvement by exploiting
natural image structures, such as learning dictionaries to encode local image
structures into sparse codes~\cite{yang_scsr_2010}. Significant
research effort has been devoted towards structure-aware approaches.
Wang {\it et al.}~\cite{wang_deep_2015} applied
LISTA (which we
discussed in Section~\ref{ssec:lista}) to patches extracted from the input
image, and recombine the predicted high-resolution patches to form a predicted
high-resolution image. A diagram depiction of the entire network architecture,
dubbed Sparsity Coding based Network (SCN), is provided in Fig.~\ref{fig:scn}. A LISTA sub-network is plugged into the end-to-end learning system, which estimates the sparse codes $\alpha$ out of all the image patches. A trainable dictionary $\bD$ then maps the sparse codes to reconstructed patches, followed by an operation which injects the patches back into the whole image. Trainable parameters of SCN include those of LISTA ($\bW_t$, $\bW_e$, $\lambda$) and the dictionary $\bD$. By
integrating the patch extraction and recombination layers into the network, the
whole network resembles a CNN as it also performs patch-by-patch
processing. The network is trained by pairing low-resolution and
high-resolution images and minimizing the Mean-Square-Error (MSE) loss~\footnote{The terminology ``MSE loss'' refers to the empirical squared loss which is an estimate of the statistical MSE.\@ We stick to this term for ease of exposition and consistency with the literature.}. In
addition to higher visual quality and PSNR gain of $0.3dB$ to $1.6dB$ over state-of-the art, the network is
faster to train and has reduced number of parameters. Fig.~\ref{fig:scn_set5} provides sample visual results from SCN and several state-of-the art techniques.

Another important application focusing on improving the quality of degraded
images is blind image deblurring. Given a sharp image blurred by an unknown
function, which is usually called the blur kernel or Point Spread Function, the
goal is to jointly estimate both the blur kernel and the underlying sharp
image. There is a wide range of approaches in the literature to blind image
deblurring: the blur kernel can be of different forms, such as Gaussian,
defocusing and motion. In addition, the blur kernel can be either spatially
uniform or non-uniform.

Blind image deblurring is a challenging topic because the blur kernel is generally
of low-pass nature, rendering the problem highly ill-posed. Most existing approaches rely on extracting stable features (such as
salient edges in natural images) to reliably estimate the blur kernels. The
sharp image can be retrieved subsequently based on the estimated kernels.
Schuler {\it et al.}~\cite{schuler_learning_2016} reviews many
existing algorithms and note that they essentially iterate over three modules:
\begin{enumerate*}
	\item feature extraction,
	\item kernel estimation and
	\item image recovery
\end{enumerate*}.

Therefore, a network can be built by unrolling and concatenating several layers of these modules, as depicted in Fig.~\ref{fig:scheuler}. More specifically, the feature extraction module is modeled as a few layers of convolutions and rectifier operations, which basically mimics a small CNN, while both the kernel and image estimation modules are modeled as least-square operations. To train the network, the blur kernels are simulated by sampling from Gaussian processes, and the blurred images are synthetically created by blurring each sharp image through a 2D discrete convolution. 

\begin{figure*}
	\centering
	\subfloat{%
		\begin{tikzpicture}[spy using outlines={rectangle,magnification=8,width=3.25cm,height=2cm}]
		\node {\includegraphics[height=0.085\textheight]{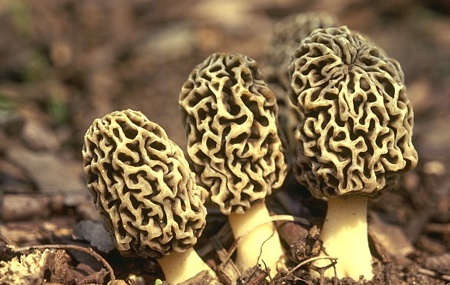}\llap{\includegraphics[height=0.03\textheight]{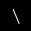}}};
		\spy[green] on (-0.62,-0.38) in node[below,ultra thick] at (0,-1.2);
		\end{tikzpicture}
	}\hspace{-3mm}
	\subfloat{%
		\begin{tikzpicture}[spy using outlines={rectangle,magnification=8,width=3.25cm,height=2cm}]
		\node {\includegraphics[height=0.085\textheight]{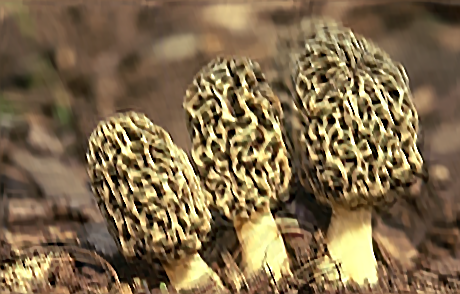}\llap{\includegraphics[height=0.03\textheight]{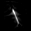}}};
		\spy[green] on (-0.62,-0.38) in node[below,ultra thick] at (0,-1.2);
		\end{tikzpicture}
	}\hspace{-3mm}
	\subfloat{%
		\begin{tikzpicture}[spy using outlines={rectangle,magnification=8,width=3.25cm,height=2cm}]
		\node {\includegraphics[height=0.085\textheight]{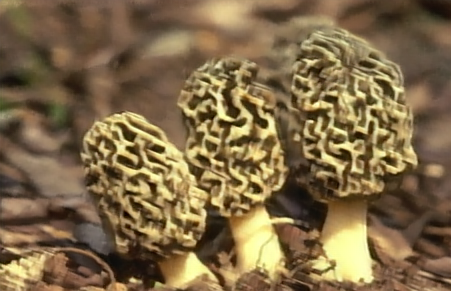}};
		\spy[green] on (-0.62,-0.38) in node[below,ultra thick] at (0,-1.2);
		\end{tikzpicture}
	}\hspace{-3mm}
	\subfloat{%
		\begin{tikzpicture}[spy using outlines={rectangle,magnification=8,width=3.25cm,height=2cm}]
		\node {\includegraphics[height=0.085\textheight]{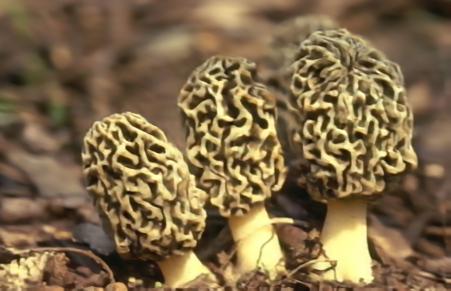}};
		\spy[green] on (-0.62,-0.38) in node[below,ultra thick] at (0,-1.2);
		\end{tikzpicture}
	}\hspace{-3mm}
	\subfloat{%
		\begin{tikzpicture}[spy using outlines={rectangle,magnification=8,width=3.25cm,height=2cm}]
		\node {\includegraphics[height=0.085\textheight]{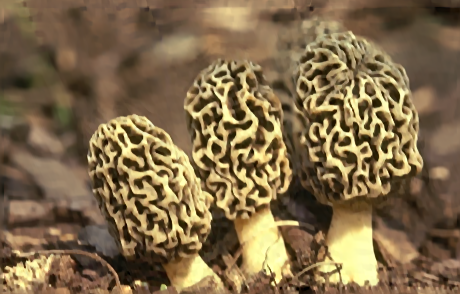}\llap{\includegraphics[height=0.03\textheight]{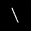}}};
		\spy[green] on (-0.62,-0.38) in node[below,ultra thick] at (0,-1.2);
		\end{tikzpicture}
	}\\
	\vspace{-3mm}
	\setcounter{subfigure}{0}
	\subfloat[Groundtruth\label{subfig:gt}]{%
		\begin{tikzpicture}[spy using outlines={rectangle,magnification=8,width=3.25cm,height=2cm}]
		\node {\includegraphics[height=0.085\textheight]{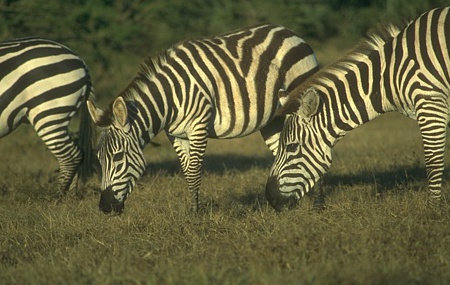}\llap{\includegraphics[height=0.03\textheight]{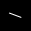}}};
		\spy[green] on (1.17,0.55) in node[below,ultra thick] at (0,-1.2);
		\end{tikzpicture}
	}\hspace{-3mm}
	\subfloat[Perrone {\it et al.}~\cite{perrone_clearer_2016}\label{subfig:perrone}]{%
		\begin{tikzpicture}[spy using outlines={rectangle,magnification=8,width=3.25cm,height=2cm}]
		\node {\includegraphics[height=0.085\textheight]{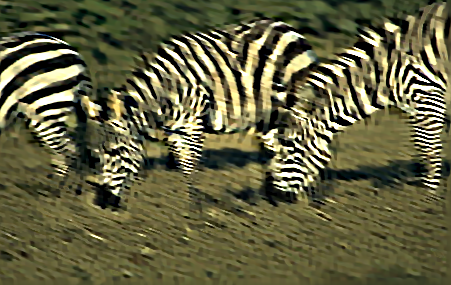}\llap{\includegraphics[height=0.03\textheight]{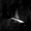}}};
		\spy[green] on (1.17,0.55) in node[below,ultra thick] at (0,-1.2);
		\end{tikzpicture}
	}\hspace{-3mm}
	\subfloat[Nah {\it et al.}~\cite{nah_deep_2017}\label{subfig:nah}]{%
		\begin{tikzpicture}[spy using outlines={rectangle,magnification=8,width=3.25cm,height=2cm}]
		\node {\includegraphics[height=0.085\textheight]{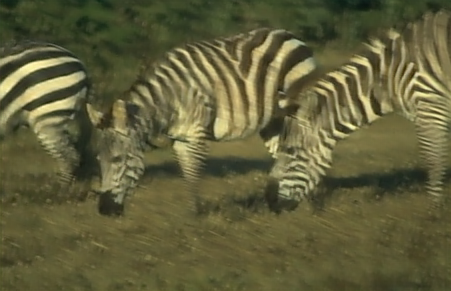}};
		\spy[green] on (1.17,0.55) in node[below,ultra thick] at (0,-1.2);
		\end{tikzpicture}
	}\hspace{-3mm}
	\subfloat[Tao {\it et al.}~\cite{tao_2018_scale}\label{subfig:tao}]{%
		\begin{tikzpicture}[spy using outlines={rectangle,magnification=8,width=3.25cm,height=2cm}]
		\node {\includegraphics[height=0.085\textheight]{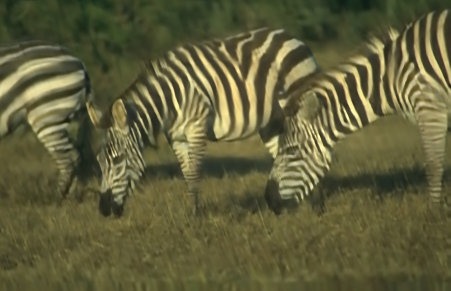}};
		\spy[green] on (1.17,0.55) in node[below,ultra thick] at (0,-1.2);
		\end{tikzpicture}
	}\hspace{-3mm}
	\subfloat[DUBLID\label{subfig:dublid}]{%
		\begin{tikzpicture}[spy using outlines={rectangle,magnification=8,width=3.25cm,height=2cm}]
		\node {\includegraphics[height=0.085\textheight]{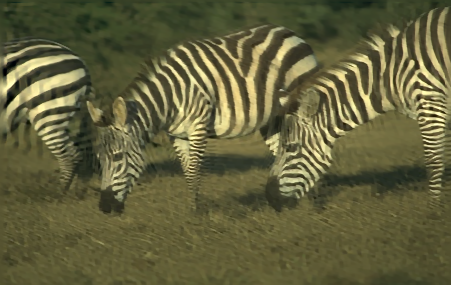}\llap{\includegraphics[height=0.03\textheight]{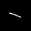}}};
		\spy[green] on (1.17,0.55) in node[below,ultra thick] at (0,-1.2);
		\end{tikzpicture}
	}
	\caption{Sample experimental results from~\cite{li_2019_deep} for visual comparison on blind image deblurring. Groundtruth images and kernels are included in \protect\subref{subfig:gt}. \protect\subref{subfig:perrone} A top-performing iterative algorithm and \protect\subref{subfig:nah}\protect\subref{subfig:tao} two state-of-the art deep learning techniques are compared against \protect\subref{subfig:dublid} the unrolling method.}\label{fig:compare_linear}
\end{figure*}

Recently, Li {\it et al.}~\cite{li_icassp19,li_2019_deep} developed an unrolling approach for blind image deblurring by enforcing sparsity constraints over filtered domains and then unrolling the half-quadratic splitting algorithm for solving the resulting optimization problem. The network is called Deep
Unrolling for Blind Deblurring (DUBLID) and detailed in the DUBLID box. As the DUBLID box reveals, custom modifications are made in the unrolled network to integrate domain knowledge that enhances the deblurring pursuit. The authors also derive custom back-propagation rules analytically to facilitate network training. Experimentally, the network offers significant performance gains and requires much fewer parameters and less inference time compared with both traditional iterative algorithms and modern neural network approaches. An example of experimental comparisons is provided in Fig.~\ref{fig:compare_linear}.

\subsection{Applications in Medical Imaging}\label{subsec:app_medical}

Medical imaging is a broad area that generally focuses on applying image
processing and pattern recognition techniques to aid clinical analysis and
disease diagnosis. Interesting topics in medical imaging include Medical
Resonance Imaging (MRI), Computed Tomography (CT) imaging, Ultrasound (US)
imaging, to name a few. Just like computational imaging, medical imaging is an
area enriched with many interesting inverse problems, and model-based
approaches such as sparse coding play a critical role in solving these
problems. In practice, data collection can be quite expensive and painstaking
for the patients, and therefore it is difficult to collect abundant samples to
train conventional deep networks. Interpretability is also an important concern. Therefore, algorithm unrolling has great
potential in this context.

In MRI, a fundamental challenge is to recover a signal from a small number of
measurements, corresponding to reduced scanning time. Yang
{\it et al.}~\cite{yang_admm_csnet} unroll the widely-known Alternating
Direction Method of Multipliers (ADMM) algorithm, a popular optimization
algorithm for solving CS and related sparsity-constrained estimation problems,
into a deep network called ADMM-CSNet. The sparsity-inducing transformations
and regularization weights are learned from real data to advance its
limited adaptability and enhance the reconstruction performance. Compared with
conventional iterative methods, ADMM-CSNet achieves the same reconstruction
accuracy using $10\%$ less sampled data and speeds up recovery by around $40$ times.
It exceeds state-of-the art deep networks by around 3dB PSNR under $20\%$
sampling rate. Refer to the box on ADMM-CSNet for further details.
\clearpage

\begin{strip}
	\begin{tcolorbox}[title={Deep Unrolling for Blind Deblurring (DUBLID)},parbox=false]
	\begin{multicols}{2}
		The spatially invariant blurring process can be represented as a discrete convolution:
		\begin{equation}
			\by=\bk\ast\bx+\bn,\label{eq:blur_model}
		\end{equation}
		where $\by$ is the blurred image, $\bx$ is the latent sharp image,
		$\bk$ is the unknown blur kernel, and $\bn$ is Gaussian random noise.
		A popular class of image deblurring algorithms perform Total-Variation (TV) minimization, which solves the following optimization problem:
		\begin{align}
			\min_{\bk,\bg_1,\bg_2}&\frac{1}{2}\left(\left\|D_x\by-\bk\ast\bg_1\right\|_2^2+\left\|D_y\by-\bk\ast\bg_2\right\|_2^2\right)\nonumber\\
								  &+\lambda_1\|\bg_1\|_1+\lambda_2\|\bg_2\|_1+\frac{\epsilon}{2}\|\bk\|_2^2,\nonumber\\
								  \text{subject to }&\|\bk\|_1=1,\quad\bk\geq 0,\label{eq:TV_min}
		\end{align} 
		where $D_x\by,D_y\by$ are the partial derivatives of $\by$ in horizontal
		and vertical directions respectively, and
		$\lambda_1,\lambda_2,\varepsilon$ are positive regularization
		coefficients. Upon convergence, the variables $\bg_1$ and $\bg_2$ are estimates of the sharp image gradients in the $x$ and $y$ directions, respectively.
		
		In~\cite{li_icassp19,li_2019_deep}~\eqref{eq:TV_min} was generalized
		by realizing that $D_x$
		and $D_y$ are computed using linear filters which can be generalized into a set of $C$ filters ${\{\mathbf{f}_i\}}_{i=1}^C$:
		\begin{align}
			\min_{\bk,{\{\bg_i\}}_{i=1}^C}&\sum_{i=1}^C\left(\frac{1}{2}\left\|\mathbf{f}_i\ast\by-\bk\ast\bg_i\right\|_2^2+\lambda_i\|\bg_i\|_1\right)+\frac{\epsilon}{2}{\|\bk\|}_2^2,\nonumber\\
			&\text{subject to }\|\bk\|_1=1,\quad\bk\geq 0.\label{eq:objective}\vspace{-6pt}
		\end{align}

		An efficient optimization algorithm to solving~\eqref{eq:objective} is the Half-Quadratic Splitting Algorithm, which alternately minimizes the surrogate problem
		\begin{align}
			\min_{\bk,{\{\bg_i,\bz_i\}}_{i=1}^C}&\sum_{i=1}^C\left(\frac{1}{2}{\left\|\mathbf{f}_i\ast\by-\bk\ast\bg_i\right\|}_2^2\right.\nonumber\\
			+&\left.\lambda_i{\|\bz_i\|}_1+\frac{1}{2\zeta_i}{\|\bg_i-\bz_i\|}_2^2\right)+\frac{\epsilon}{2}{\|\bk\|}_2^2,\nonumber\\
			\text{subject to }&\|\bk\|_1=1,\quad\bk\geq 0,\label{eq:penalty}
		\end{align}
		over the variables ${\{\bg_i\}}_{i=1}^C$, ${\{\bz_i\}}_{i=1}^C$ and $\bk$ sequentially. Here $\zeta_i,i=1,\dots,C$ are regularization coefficients. A noteworthy fact is that each individual minimization admits an analytical expression, which facilitates casting~\eqref{eq:penalty} into network layers. Specifically, in the $l$-th iteration ($l\geq 0$) the following updates are performed:
		\begin{align}
			\bg_i^{l+1}&=\cF^{-1}\left\{\frac{\zeta^l_i\widehat{\bk^l}^\ast\odot\widehat{\mathbf{f}^l_i}\odot\widehat{\by}+\widehat{\bz_i^l}}{\zeta^l_i\left|\widehat{\bk^l}\right|^2+1}\right\}\nonumber\\
					   &:=\cM^1\left\{\mathbf{f}^l\ast\by,\bz^l;\zeta^l\right\},\quad\forall i,\nonumber\\
			\bz_i^{l+1}&=\cS_{\lambda^l_i\zeta^l_i}\left\{\bg_i^{l+1}\right\}\nonumber\\
			&:=\cM^2\left\{\bg^{l+1};\beta^l\right\},\quad\forall i\label{eq:analytic_expressions}\\
			\bk^{l+1}&=\cN_1{\left[\cF^{-1}\left\{\frac{\sum_{i=1}^C\widehat{\bz_i^{l+1}}^\ast\odot\widehat{\mathbf{f}^l_i}\odot\widehat{\by_i}}{\sum_{i=1}^C\left|\widehat{\bz_i^{l+1}}\right|^2+\epsilon}\right\}\right]}_+\nonumber\\
					 &:=\cM^3\left\{\mathbf{f}^l\ast\by,\bz^{l+1}\right\},\nonumber
		\end{align}
		where ${[\cdot]}_+$ is the ReLU operator, $\widehat{\bx}$ denotes the Discrete Fourier Transform (DFT) of $\bx$, $\cF^{-1}$ indicates the inverse DFT operator, $\odot$ refers to elementwise multiplication, $\cS$ is the soft-thresholding operator defined elementwise in~\eqref{eq:soft_thresh}, and the operator $\cN_1(\cdot)$ normalizes its operand into unit sum. Here $\zeta^l={\{\zeta^l_i\}}_{i=1}^C$, $\beta^l={\{\lambda^l_i\zeta^l_i\}}_{i=1}^C$, and $\bg^l$, $\mathbf{f}^l\ast\by$, $\bz^l$ refer to ${\{\bg^l_i\}}_{i=1}^C$, ${\{\mathbf{f}_i^l\ast\by\}}_{i=1}^C$, ${\{\bz^l_i\}}_{i=1}^C$ stacked together. Note that layer-specific parameters $\zeta^l,\beta^l,\mathbf{f}^l$ are used. The parameter $\epsilon > 0$ is a fixed constant.
		
		As with most existing unrolling methods, only $L$ iterations are performed. The sharp image is retrieved from $\bg^L$ and $\bk^L$ by solving the following linear least-squares problem:
		\begin{align}
			\widetilde{\bx}&=\argmin_{\bx}\frac{1}{2}\left\|\by-\widetilde{\bk}\ast\bx\right\|_2^2+\sum_{i=1}^C\frac{\eta_i}{2}\left\|{\mathbf{f}}^L_i\ast\bx-\bg^L_i\right\|_2^2\nonumber\\
						   &=\cF^{-1}\left\{\frac{\widehat{\widetilde{\bk}}^\ast\odot\widehat{\by}+\sum_{i=1}^C\eta_i\widehat{{\mathbf{f}}^L_i}^\ast\odot\widehat{\bg^L_i}}{\left|\widehat{\widetilde{\bk}}\right|^2+\sum_{i=1}^C\eta_i\left|\widehat{{\mathbf{f}}^L_i}\right|^2}\right\}\nonumber\\
						   &:=\cM^4\left\{\by,\bg^L,\bk^L;\eta,\mathbf{f}^L\right\},\label{eq:image_retrieval}
		\end{align}
		where $\mathbf{f}^L={\{\mathbf{f}_i^L\}}_{i=1}^C$ are the filter coefficients in the $L$-th layer, and $\eta={\{\eta_i\}}_{i=1}^C$ are positive regularization coefficients. By unrolling~\eqref{eq:analytic_expressions} and~\eqref{eq:image_retrieval} into a deep network, we get $L$ layers of $\bg$, $\bz$ and $\bk$ updates, followed by one layer of image retrieval. The filter coefficients ${\mathbf{f}^l_i}$'s and regularization parameters $\{\lambda^l_i,\zeta^l_i,\eta_i\}$'s are learned by back-propagation. Note that $\mathbf{f}_i^L$'s are shared in both~\eqref{eq:analytic_expressions} and~\eqref{eq:image_retrieval} and are updated jointly. The final network architecture is depicted in Fig.~\ref{fig:dublid}.

		Similar to~\cite{schuler_learning_2016}, the network is trained using synthetic samples, i.e., by convolving the sharp images to obtain blurred versions. The training loss function is the translation-invariant MSE loss to compensate for the possible spatial shifts of the deblurred images and the blur kernel.
	\end{multicols}
\end{tcolorbox}
\end{strip}

\begin{figure*}[h!]
	\includegraphics[width=\textwidth]{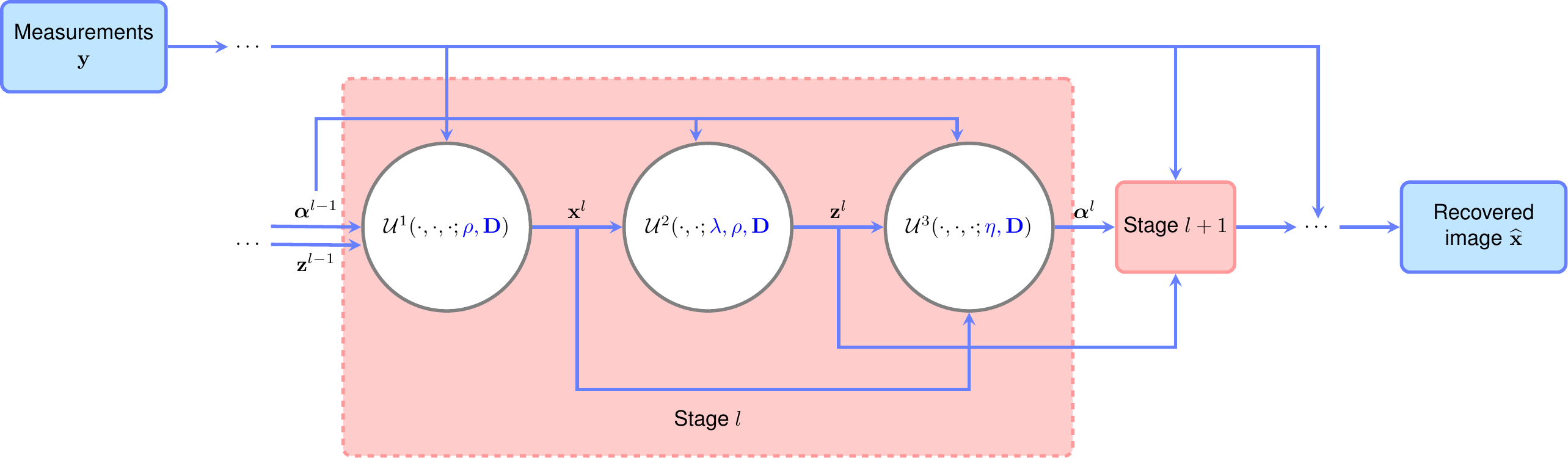}
	\caption{Diagram representation of ADMM-CSNet~\cite{yang_admm_csnet}: each stage comprises a series of inter-related operations, whose analytic forms are given in~\eqref{eq:admm_csnet_update}. The trainable parameters are colored in blue.}\label{fig:unroll_admm_csnet}
\end{figure*}

\begin{figure*}
	\centering
	\includegraphics[width=\textwidth]{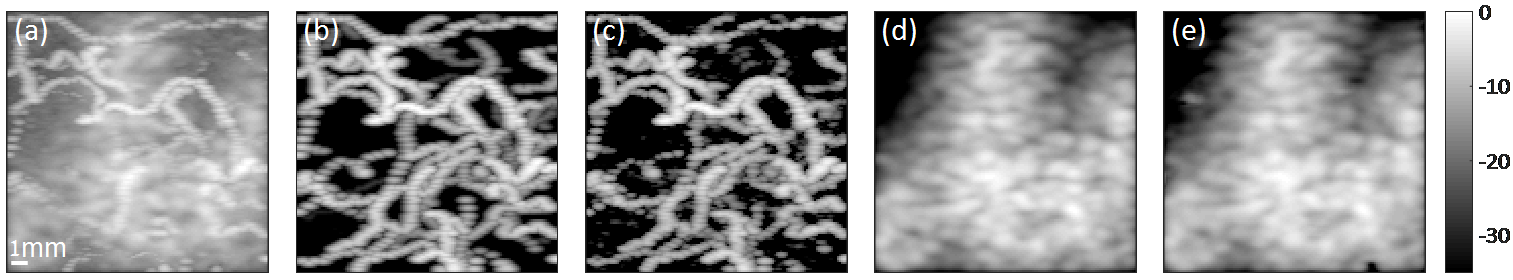}
	\caption{Sample experimental results demonstrating recovery of Ultrasound Contrast Agents (UCAs) from cluttered Maximum Intensity Projection (MIP) images~\cite{solomon_deep_2018}. (a) MIP image of the input movie, composed from 50 frames of simulated UCAs
		cluttered by tissue. (b) Ground-truth UCA MIP image. (c) Recovered UCA MIP image via CORONA.\@ (d) Ground-truth tissue
	MIP image. (e) Recovered tissue MIP image via CORONA.\@ Color bar is in dB. Figure reproduced from~\cite{solomon_deep_2018} with authors' permission.}\label{fig:corona}
\end{figure*}

As another work on MRI reconstruction, Hosseini {\it et
	al.}~\cite{hosseini_dense_2019} unroll the well-known Proximal Gradient
	Descent (PGD) algorithm into a deep network. Motivated by
	momentum-based acceleration techniques such as Nesterov's
	method~\cite{nesterov_gradient_2013}, they introduce dense connections into
	their network, which facilitate information flow across non-adjacent
	layers. Performance improvements over conventional PGD-based methods are
shown experimentally.

In tomographic reconstruction, Adler {\it et al.}~\cite{adler2018learned}
	unroll the Primal Dual Hybrid Gradient (PDHG) algorithm, a well-known
	technique for primal-dual non-smooth optimization. They substitute the primal and dual proximal operators with certain parameterized operators
	such as CNNs, and train both the operator parameters and the algorithm
	parameters in an end-to-end fashion. Their method demonstrates 
	improvements over conventional approaches in recovering low dose CT images.\@ While this
	technique offers merits in reconstruction, the extracted
	features may not favor
	detection
	tasks. Therefore, Wu {\it et al.}~\cite{wu_end--end_2018} extend their
	method
	by concatenating it with a detection network, and apply joint fine-tuning after
	training both networks individually. Their jointly fine-tuned network outperforms
state of the art alternatives.

\begin{strip}
\begin{tcolorbox}[title={ADMM-CSNet},parbox=false]
	\begin{multicols}{2}
		Consider random linear measurements $\by\in\mathbb{C}^m$ formed by $\by\approx\bphi\bx$ where $\bphi\in\mathbb{C}^{m\times n}$ is a measurement matrix with $m<n$, Compressive Sensing (CS) aims at reconstructing the original signal $\bx\in\mathbb{R}^n$ by exploiting its underlying sparse structure in a transform domain~\cite{eldar_2012_compressed}.

		A generalized CS model can be formulated as the following optimization problem~\cite{yang_admm_csnet}:
		\begin{align}
			\min_{\bx}\frac{1}{2}\|\bphi\bx-\by\|_2^2+\sum_{i=1}^C\lambda_i g(\bD_i\bx),\label{eq:general_cs}
		\end{align}
		where $\lambda_i$'s are positive regularization coefficients, $g(\cdot)$ is a sparsity-inducing function, and ${\{\bD_i\}}_{i=1}^C$ is a sequence of $C$ operators, which effectively perform linear filtering operations. Concretely, $\bD_i$ can be taken as a wavelet transform and $g$ can be chosen as the $\ell^1$ norm. However, for better performance the method in~\cite{yang_admm_csnet} learns both of them from an unrolled network.

		An efficient minimization algorithm for solving~\eqref{eq:general_cs} is the Alternating Direction Method of Multipliers (ADMM)~\cite{boyd_2011_distributed}. Problem~\eqref{eq:general_cs} is first recast into a constrained minimization through variable splitting:
		\begin{align}
			\min_{\bx,{\{\bz\}}_{i=1}^C}&\frac{1}{2}\|\bphi\bx-\by\|_2^2+\sum_{i=1}^C\lambda_i g(\bz_i),\nonumber\\
			\text{subject to }&\bz_i=\bD_i\bx,\quad\forall i.\label{eq:split_cs}
		\end{align}
		The corresponding augmented Lagrangian is then formed as follow:
		\begin{align}
			\cL_\rho(\bx,\bz;\balpha_i)&=\frac{1}{2}\|\bphi\bx-\by\|_2^2+\sum_{i=1}^C\lambda_i g(\bz_i)\nonumber\\
									 &+\frac{\rho_i}{2}\|\bD_i\bx-\bz_i+\balpha_i\|_2^2,\label{eq:aug_lagrangian}
		\end{align}
		where ${\{\balpha_i\}}_{i=1}^C$ are dual variables and ${\{\rho_i\}}_{i=1}^C$ are penalty coefficients. ADMM then alternately minimizes~\eqref{eq:aug_lagrangian} followed by a dual variable update, leading to the following iterations:
		\begin{align}
			\bx^l&={\left(\bphi^H\bphi+\sum_{i=1}^C\rho_i\bD_i^T\bD_i\right)}^{-1}\bigl[\bphi^H\by\nonumber\\
			&+\sum_{i=1}^C \rho_i {\bD_i}^T\left({\bz_i}^{l-1}-{\balpha_i}^{l-1}\right)\bigr]\nonumber\\
			&:=\cU^1\left\{\by,\balpha_i^{l-1},\bz_i^{l-1};\rho_i,\bD_i\right\},\nonumber\\
			\bz_i^l&=\cP_{g}\left\{\bD_i\bx^l+\balpha_i^{l-1};\frac{\lambda_i}{\rho_i}\right\}\label{eq:admm_csnet_update}\\
			&:=\cU^2\left\{\balpha_i^{l-1},\bx^l;\lambda_i,\rho_i,\bD_i\right\},\nonumber\\
			\balpha_i^l&=\balpha_i^{l-1}+\eta_i(\bD_i\bx^l-\bz^l_i)\nonumber\\
					   &:=\cU^3\left\{\balpha_i^{l-1},\bx^l,\bz_i^l;\eta_i,\bD_i\right\},\quad\forall i,\nonumber
		\end{align}
		where $\eta_i$'s are constant parameters, and $\cP_g\{\cdot;\lambda\}$ is the proximal mapping for $g$ with parameter $\lambda$. The unrolled network can thus be constructed by concatenating these operations and learning the parameters $\lambda_i,\rho_i,\eta_i,\bD_i$ in each layer. Fig.~\ref{fig:unroll_admm_csnet} depicts the resulting unrolled network architecture. In~\cite{yang_admm_csnet} the authors discuss several implementation issues, including efficient matrix inversion and the back-propagation rules. The network is trained by minimizing a normalized version of the Root-Mean-Square-Error.
	\end{multicols}

\end{tcolorbox}
\end{strip}

Another important imaging modality is ultrasound, which has the advantage of
being a radiation-free approach.  When used for blood flow depiction, one of
the challenges is the fact that the tissue reflections tend to be much stronger
than those of the blood, leading to strong clutter resulting from the tissue. 
Thus, an important task is to separate the tissue from the blood. Various
filtering methods have been used in this context such as high pass filtering,
and filtering based on the singular value decomposition.  Solomon et
al.~\cite{solomon_deep_2018} suggest using a robust Principal Component
Analysis (PCA) approach by modeling the received ultrasound movie as a low-rank
and sparse matrix where the tissue is low rank and the blood vessels are
sparse.  They then unroll an ISTA approach to robust PCA into a deep network,
called Convolutional rObust pRincipal cOmpoNent Analysis (CORONA). As the name
suggests, they replace matrix multiplications with convolutional layers,
effectively converting the network into a CNN-like architecture. Compared with
state-of-the-art approaches, CORONA demonstrates vastly improved
reconstruction quality and has much fewer parameters than the well-known
ResNet~\cite{he_deep_2016}. Refer to the box on Convolutional rObust pRincipal cOmpoNent Analysis for details. LISTA-based methods have also been applied in ultrasound to improve image super-resolution~\cite{dardikman-yoffe_learned_2020}.

\subsection{Applications in Vision and Recognition}\label{subsec:app_vision}

Computer vision is a broad and fast growing area that has achieved tremendous
success in many interesting topics in recent years. A major driving force for
its rapid progress is deep learning. For instance, thanks to the availability
of large scale training samples, in image recognition tasks researchers have
surpassed human-level performance over the ImageNet dataset by employing deep
CNN~\cite{he_delving_2015}. Nevertheless, most existing approaches to date are
highly empirical, and lack of interpretability has become an increasingly
serious issue. To overcome this drawback, recently researchers are paying more
attention to algorithm unrolling~\cite{zheng_conditional_2015,liu_deep_2018}.

One example is in semantic image segmentation, which assigns class labels to each
pixel in an image. Compared with traditional low-level image segmentation, it provides
additional information about object categories, and thus creates semantically
meaningful segmented objects. By performing pixel-level labeling, semantic
segmentation can also be regarded as an extension to image recognition.
Applications of semantic segmentation include autonomous driving, robot vision,
and medical imaging.

Traditionally Conditional Random Field (CRF) was a popular approach.
Recently deep networks have become the primary tool. Early works of deep
learning approaches are capable of recognizing objects at a high level;
however, they are relatively less accurate in delineating the objects than
CRFs.

Zheng {\it et al.\/}~\cite{zheng_conditional_2015} unroll the Mean-Field (MF) iterations of CRF into a RNN, and then concatenate the semantic segmentation network with this RNN to form a deep network. The concatenated network resembles conventional semantic segmentation followed by CRF-based post-processing, while end-to-end training can be performed over the whole network. Liu {\it et al.\/}~\cite{liu_deep_2018} follow the same direction to construct their segmentation network, called Deep Parsing Network. In their approach, they adopt a generalized pairwise energy and perform MF iteration only once for the purposes of efficiency. Refer to the box on Unrolling CRF into RNN for further details.

\subsection{Other Signal Processing Applications}\label{subsec:app_speech}
Until now, we have surveyed compelling applications in image processing and computer vision. Algorithm unrolling has also been successfully applied to a variety of other signal processing domains. We next consider speech processing, which is one of the fundamental problems in digital signal processing. Topics in speech processing include recognition, coding, synthesis, and more. Among all problems of interest, source separation stands out as a challenging yet intriguing one. Applications of source separation include speech enhancement and recognition.

For single channel speech separation, Non-negative Matrix Factorization (NMF) is a widely-applied technique. Recently,
Hershey {\it et al.}~\cite{hershey_2014_deep} unrolled NMF into a deep network, dubbed \emph{deep NMF}, as a concrete realization of their abstract unrolling framework. Detailed descriptions are in the box Deep NMF.\@ The deep NMF was evaluated on the task of speech enhancement in reverberated noisy mixtures, using a dataset collected from the Wall Street Journal. Deep NMF was shown to outperform both a conventional deep neural network~\cite{hershey_2014_deep} and the iterative sparse NMF method~\cite{eggert_2004_sparse}.

Wang {\it et al.}~\cite{wang_2018_end} propose an end-to-end training approach
for speech separation, by casting the commonly employed forward and inverse
Short-time Fourier Transform (STFT) operations into network layers, and
concatenate them with an iterative phase reconstruction algorithm, dubbed
Multiple Input Spectrogram Inversion~\cite{guanwan_2010_iterative}. In
doing so, the loss function acts on reconstructed signals rather than their
STFT magnitudes, and phase inconsistency can be reduced through training. The
trained network exhibits $1$ dB higher SNR
over state-of-the art techniques on public datasets.

Monitoring the operating conditions of power grids in real time is a critical task when deploying large-scale contemporary power grids. To address the computational complexity issue of conventional power system state estimation methods, Zhang {\it et al.}~\cite{zhang_real-time_2019} unroll an iterative physics-based prox-linear solver into a deep neural network. They further extend their approach for state forecasting. Numerical experiments on the IEEE 57-and 118-bus benchmark systems confirm its improved performance over alternative approaches.

Multi-spectral image fusion is a fundamental problem in remote sensing.
Lohit {\it et al.}~\cite{lohit_unrolled_2019} unroll the projected gradient
descent algorithm for fusing low spatial resolution multi-spectral aerial
images with their associated high resolution panchromatic counterpart. They
also show experimental improvements over several baselines.

Finally, unrolling has also been applied to super resolution microscopy~\cite{dardikman-yoffe_learned_2020}. Here the authors unroll the Sparsity-based Super-resolution Microscopy from Correlation Information (SPARCOM) method~\cite{solomon_sparsity-based_2018} which performs sparse recovery in the correlation domain.

\begin{strip}
\begin{tcolorbox}[title={Convolutional Robust Principal Component Analysis},parbox=false]
	\begin{multicols}{2}
		In US imaging, a series of pulses are transmitted into the imaged medium, and their echoes are received in each transducer element. After beamforming and demodulation, a series of movie frames are acquired. Stacking them together as column vectors leads to a data matrix $\bD\in\mathbb{C}^{m\times n}$, which can be modeled as follows:
		\begin{align*}
			\bD=\bH_1\bL+\bH_2\bS+\bN,
		\end{align*}
		where $\bL$ comprises the tissue signals, $\bS$ comprises the echoes returned from the blood signals, $\bH_1,\bH_2$ are measurement matrices, and $\bN$ is the noise matrix. Due to its high spatial-temporal coherence, $\bL$ is typically a low-rank matrix, while $\bS$ is generally a sparse matrix since blood vessels usually sparsely populate the imaged medium.

		Based on these observations, the echoes $\bS$ can be estimated through a transformed low-rank and sparse decomposition by solving the following optimization problem:
		\begin{align}
			\min_{\bL,\bS}\frac{1}{2}\|\bD-(\bH_1\bL+\bH_2\bS)\|_F^2+\lambda_1\|\bL\|_\ast+\lambda_2\|\bS\|_{1,2},\label{eq:low_rank_sparse_decomp}
		\end{align}
		where $\|\cdot\|_\ast$ is the nuclear norm of a matrix which promotes low-rank solutions, and $\|\cdot\|_{1,2}$ is the mixed $\ell_{1,2}$ norm which enforces row sparsity. Problem~\eqref{eq:low_rank_sparse_decomp} can be solved using a generalized version of ISTA in the matrix domain, by utilizing the proximal mapping corresponding to the nuclear norm and mixed $\ell_{1,2}$ norm. In the $l$-th iteration it executes the following steps:
		\begin{align*}
			\bL^{l+1}&=\cT_{\frac{\lambda_1}{\mu}}\left\{\left(\bI-\frac{1}{\mu}\bH_1^H\bH_1\right)\bL^l-\bH_1^H\bH_2\bS^l+\bH_1^H\bD\right\},\\
		\end{align*}
		\begin{align*}
			\bS^{l+1}&=\cS^{1,2}_{\frac{\lambda_2}{\mu}}\left\{\left(\bI-\frac{1}{\mu}\bH_2^H\bH_2\right)\bS^l-\bH_2^H\bH_1\bL^l+\bH_2^H\bD\right\},
		\end{align*}
		where $\cT_\lambda\{\bX\}$ is the singular value thresholding operator
		that performs soft-thresholding over the singular values of $\bX$ with
		threshold $\lambda$, $\cS^{1,2}_\lambda$ performs row-wise
		soft-thresholding with parameter $\lambda$, and $\mu$ is the step size
		parameter for ISTA.\@ Technically, $\cT_\lambda$ and $\cS^{1,2}_\lambda$
		correspond to the proximal mapping for the nuclear norm and mixed
		$\ell^{1,2}$ norm, respectively. Just like the migration from MLP to CNN, the matrix multiplications can be replaced by convolutions, which gives rise to the following iteration steps instead:
		\begin{align}
			\bL^{l+1}&=\cT_{\lambda_1^l}\left\{\bP_5^l\ast\bL^l+\bP_3^l\ast\bS^l+\bP_1^l\ast\bD\right\},\label{eq:corona_lupdate}\\
			\bS^{l+1}&=\cS^{1,2}_{\lambda_2^l}\left\{\bP_6^l\ast\bS^l+\bP_4^l\ast\bL^l+\bP_2^l\ast\bD\right\},\label{eq:corona_supdate}
		\end{align}
	where $\ast$ is the convolution operator. Here $\bP^l_i,i=1,\dots,6$ are a series of convolution filters that are learned from the data in the $l$-th layer, and $\lambda_1^l,\lambda_2^l$ are thresholding parameters for the $l$-th layer. By casting~\eqref{eq:corona_lupdate} and~\eqref{eq:corona_supdate} into network layers, a deep network resembling CNN is formed. The parameters $\bP^l_i,i=1,2,\ldots,6$ and $\{\lambda_1^l,\lambda_2^l\}$ are learned from training data.

		To train the network, one can first obtain ground-truth $\bL$ and $\bS$ from $\bD$ by executing ISTA-like algorithms up to convergence. Simulated samples can also be added to address lack of training samples. MSE losses are imposed on $\bL$ and $\bS$, respectively.
	\end{multicols}
\end{tcolorbox}
\end{strip}

\subsection{Enhancing Efficiency Through Unrolling}\label{subsec:app_efficiency}
In addition to intepretability and performance improvements, unrolling can provide significant advantages for practical deployment, including higher computational efficiency and lower number of parameters, which in turn lead to reduced memory footprints and storage requirements. Table~\ref{tab:efficiency} summarizes selected results from recent unrolling papers to illustrate such benefits. For comparison, results for one iterative algorithm and one deep network are included, both selected from representative top-performing methods. Note further, that for any two methods compared in Table~\ref{tab:efficiency}, the run times are reported on consistent implementation platforms. More details can be found in the respective papers.

\begin{figure*}
	\includegraphics[width=\textwidth]{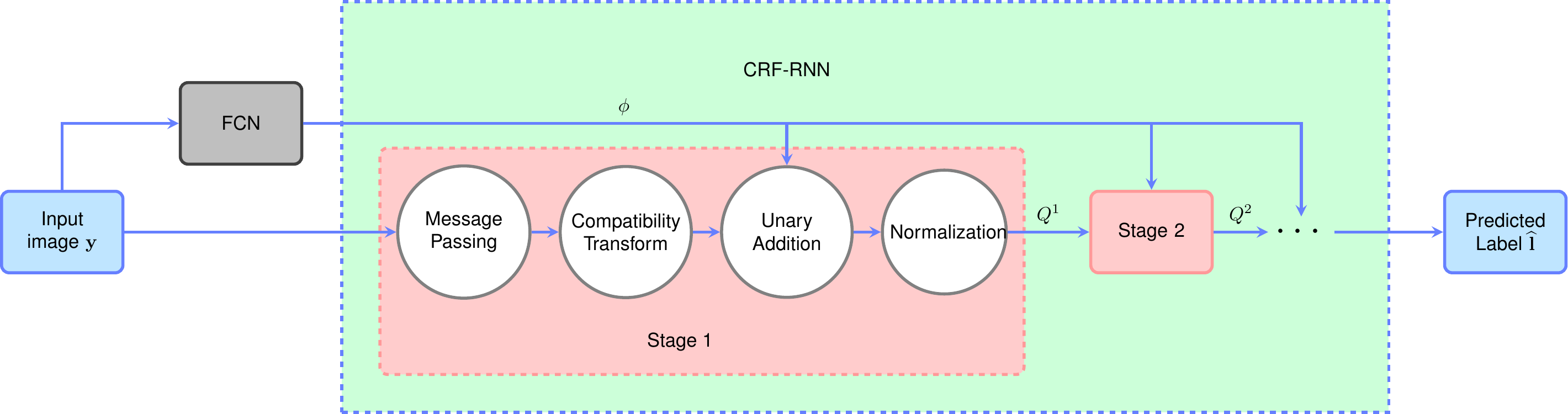}
	\caption{Diagram representation of the CRF-RNN network~\cite{zheng_conditional_2015}: a FCN is concatenated with a RNN called CRF-RNN to form a deep network. This RNN essentially performs MF iterations and acts like CRF-based post-processing. The concatenated network can be trained end-to-end to optimize its performance.}\label{fig:dpn}
\end{figure*}

Compared to their iterative counterparts, unrolling often dramatically boosts the computational speed. For instance, it was reported in~\cite{lecun_efficient_2012} that LISTA may be $20$ times faster than ISTA after training; DUBLID~\cite{li_2019_deep} can be $1000$ times faster than TV-based deblurring; ADMM-CSNet~\cite{yang_admm_csnet} can be about four times faster than the BM3D-AMP algorithm~\cite{metzler_2016_denoising}; CORONA~\cite{solomon_deep_2018} is over $50$ times faster than the Fast-ISTA algorithm; the prox-linear network proposed by Zhang~\cite{zhang_real-time_2019} is over $500$ faster than the Guass-Newton algorithm.

Typically, by embedding domain-specific structures into the network, the unrolled networks need much fewer parameters than conventional networks that are less specific towards particular applications. For instance, the number of parameters for DUBLID is more than $100$ times lower than SRN~\cite{tao_2018_scale}, while CORONA~\cite{solomon_deep_2018} has an order of magnitude lower number of parameters than ResNet~\cite{he_deep_2016}.

Under circumstances where each iteration (layer) can be executed highly efficiently, the unrolled networks may even be more efficient than conventional networks. For instance, ADMM-CSNet~\cite{yang_admm_csnet} has shown to be about twice as fast as ReconNet~\cite{kulkarni_2016_reconnet}, while DUBLID~\cite{liu_deep_2018} is almost two times faster than DeblurGAN~\cite{kupyn_deblurgan_2018}.

\begin{strip}
\begin{tcolorbox}[title={Unrolling CRF into RNN},parbox=false]
	\begin{multicols}{2}
		CRF is a fundamental model for labeling undirected graphical models. A special case of CRF, where only pairwise interactions of graph nodes are considered, is the Markov random field. Given a graph $(\cV,\cE)$ and a predefined collection of labels $\cL$, it assigns label $l_\bp\in\cL$ to each node $\bp$ by minimizing the following energy function:
		\begin{align*}
			E({\{l_\bp\}}_{\bp\in\cV})=\sum_{\bp\in\cV}\phi_\bp(l_\bp)+\sum_{(\bp,\bq)\in\cE}\psi_{\bp,\bq}(l_\bp,l_\bq),
		\end{align*}
		where $\phi_\bp(\cdot)$ and $\psi_{\bp,\bq}(\cdot)$ are commonly called unary energy and pairwise energy, respectively. Typically, $\phi_\bp$ models the preference of assigning $\bp$ with each label given the observed data, while $\psi_{\bp,\bq}$ models the smoothness between $\bp$ and $\bq$. In semantic segmentation, $\cV$ comprises the image pixels, $\cE$ is the set of pixel pairs, while $\cL$ consists of object categories, respectively.

		In~\cite{zheng_conditional_2015}, the unary energy $\phi_\bp$ is chosen as the output of a semantic segmentation network, such as the well-known Fully Convolutional Network (FCN)~\cite{long_fully_2015}, while the pairwise energy $\psi_{\bp,\bq}(\mathbf{f}_\bp,\mathbf{f}_\bq)$ admits the following special form:
		\begin{align*}
			\psi(l_\bp,l_\bq)=\mu(\bp,\bq)\sum_{m=1}^M w^m G^m(\mathbf{f}_\bp,\mathbf{f}_\bq),
		\end{align*}
		where ${\{G^m\}}_{m=1}^M$ is a collection of Gaussian kernels and ${\{w^m\}}_{m=1}^M$ are the corresponding weights. Here $\mathbf{f}_\bp$ and $\mathbf{f}_\bq$ are the feature vectors for pixel $\bp$ and $\bq$, respectively, and $\mu(\cdot,\cdot)$ models label compatibility between pixel pairs.

		An efficient inference algorithm for energy minimization over fully-connected CRFs is the MF iteration~\cite{krahenbuhl_2011_efficient}, which executes the following steps iteratively:
		\begin{align}
			&(\text{Message Passing}):\tilde{Q}_\bp^m(l)\gets\sum_{j\neq i}G^m(\mathbf{f}_i,\mathbf{f}_j)Q_\bq(l),\nonumber\\
			&(\text{Compatibility Transform}):\\
			&\qquad\hat{Q}_\bp(l_\bp)\gets\sum_{l\in\cL}\sum_m \mu^m(l_\bp,l)w^m\tilde{Q}_\bp^m(l),\label{eq:mean_field}\\
			&(\text{Unary Addition}):Q_\bp(l_\bp)\gets\exp\left\{-\phi_\bp(l_\bp)-\hat{Q}_\bp(l_\bp)\right\},\nonumber\\
			&(\text{Normalization}):Q_\bp(l_\bp)\gets\frac{Q_\bp(l_\bp)}{\sum_{l\in\cL}Q_\bp(l)},\nonumber
		\end{align}
		where $Q_\bp(l_\bp)$ can be interpreted as the margin probability of assigning $\bp$ with label $l_\bp$. A noteworthy fact is that each update step resembles common neural network layers. For instance, message passing can be implemented by filtering through Gaussian kernels, which imitates passing through a convolutional layer. The compatibility transform can be implemented through $1\times 1$ convolution, while the normalization can be considered as the popular soft-max layer. These layers can thus be unrolled to form a RNN, dubbed CRF-RNN.\@ By concatenating FCN with it, a network which can be trained end-to-end is formed. A diagram illustration is in presented Fig.~\ref{fig:dpn}.
	\end{multicols}
\end{tcolorbox}
\end{strip}

\begin{figure*}
\includegraphics[width=\textwidth]{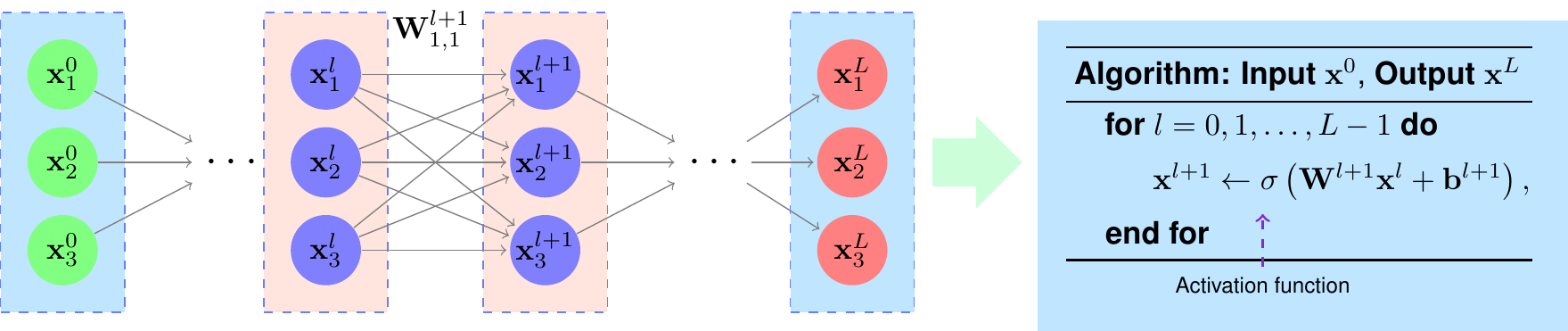}
	\caption{A MLP can be interpreted as executing an underlying iterative algorithm with finite iterations and layer-specific parameters.}\label{fig:reinterp}
\end{figure*}

\begin{strip}
	\begin{tcolorbox}[title={Deep (Unrolled) NMF},parbox=false,label=box:dnmf]
		\begin{multicols}{2}
			Single channel source separation refers to the task of decoupling several source signals from their mixture. Suppose we collect a sequence of $T$ mixture frames, where $\mathbf{m}_t\in\mathbb{R}^F_+,t=1,2,\dots,T$ is the $t$-th frame. Given a set of non-negative basis vectors ${\{\bw_l\in\mathbb{R}^F_+\}}_{l=1}^L$, we can represent $\mathbf{m}_t$ (approximately) by
			\begin{equation}
				\mathbf{m}_t\approx\sum_{l=1}^L\bw_l h_{lt},\label{eq:nonneg_repr}
			\end{equation}
			where $h_{lt}$'s are the coefficients chosen to be non-negative. By stacking $\mathbf{m}_t$'s column by column, we form a non-negative matrix $\bM\in\mathbb{R}_+^{F\times T}$, so that~\eqref{eq:nonneg_repr} can be expressed in matrix form:
			\begin{align}
				\bM\approx\bW\bH,\qquad\bW\geq 0,\bH\geq 0\label{eq:nonneg_fact}
			\end{align}
		where $\bW$ has $\bw_l$ as its $l$-th column, $\geq 0$ denotes elementwise non-negativity, and $\bH={(h_{l,t})}$. To remove multiplicative ambiguity, it is commonly assumed that each column of $\bW$ has unit $\ell^2$ norm, i.e., $\bw_l$'s are unit vectors. The model~\eqref{eq:nonneg_fact} is commonly called Non-negative Matrix Factorization (NMF)~\cite{lee_1999_learning} and has found wide applications in signal and image processing. In practice, the non-negativity constraints prevent mutual cancelling of basis vectors and thus encourage semantically meaningful decompositions, which turns out highly beneficial.

		Assuming the phases among different sources are approximately the same, the power or magnitude spectrogram of the mixture can be decomposed as a summation of those from each sources. Therefore, after performing NMF, the sources can be separated by selecting basis vectors corresponding to each individual source and recombining the source-specific basis vectors to recover the magnitude spectrograms. In practical implementation, typically a filtering process similar to the classical Wiener filtering is performed for magnitude spectrogram recovery.

			To determine $\bW$ and $\bH$ from $\bM$, one may consider solving the following optimization problem~\cite{fevotte_2009_nonnegative}:
			\begin{align}
				\widehat{\bW},\widehat{\bH}&=\arg\min_{\bW\geq 0,\bH\geq 0}D_\beta\left(\bM|\bW\bH\right) + \mu\|\bH\|_1,\label{eq:nmf_obj}
			\end{align}
			where $D_\beta$ is the $\beta$-divergence, which can be considered as a generalization of the well-known Kullback-Leibler divergence, and $\mu$ is a regularization parameter that controls the sparsity of the coefficient matrix $\bH$. By employing a majorization-minimization scheme, problem~\eqref{eq:nmf_obj} can be solved by the following multiplicative updates:
			\begin{align}
				&\bH^l=\bH^{l-1}\odot\frac{\bW^T\left[\bM\odot{\left(\bW\bH^{l-1}\right)}^{\beta-2}\right]}{\bW^T{\left(\bW\bH^{l-1}\right)}^{\beta-1}+\mu},\label{eq:hupdate}\\
				&\bW^l=\bW^{l-1}\odot\frac{\left[\bM\odot{\left(\bW^{l-1}\bH^l\right)}^{\beta-2}\right]{\bH^l}^T}{{\left(\bW^{l-1}\bH^l\right)}^{\beta-1}{\bH^l}^T},\label{eq:wupdate}\\
				&\text{Normalize } \bW^l \text{ so that the columns of }\bW^l\nonumber\\
				&\text{ have unit norm and scale }\bH^l\text{ accordingly},\label{eq:normalize}
			\end{align}
			for $l=1,2,\dots$. In~\cite{hershey_2014_deep}, a slightly different update scheme for $\bW$ was employed to encourage the discriminative power. We omit discussing it for brevity.
			
			A deep network can be formed by unfolding these iterative updates. In~\cite{hershey_2014_deep}, $\bW^l$'s are untied from the update rule~\eqref{eq:wupdate} and considered trainable parameters. In other words, only~\eqref{eq:hupdate} and~\eqref{eq:normalize} are executed in each layer. Similar to~\eqref{eq:nmf_obj}, the $\beta$-divergence, with a different $\beta$ value, was employed in the training loss function. A splitting scheme was also designed to preserve the non-negativity of $\bW^l$'s during training.
		\end{multicols}
	\end{tcolorbox}
\end{strip}

\section{Conceptual Connections and Theoretical Analysis}\label{sec:connections}
In addition to creating efficient and interpretable network architectures which
achieve superior performance in practical applications, algorithm unrolling can
provide valuable insights from a conceptual standpoint. As detailed in the previous section, solutions to real-world signal processing problems often exploit domain specific prior knowledge. Inheriting this domain
knowledge is of both conceptual and practical importance in deep learning
research. To this end, algorithm unrolling can potentially serve as a
powerful tool to help establish conceptual connections between prior-information guided analytical methods and modern neural networks.

In particular, algorithm unrolling may be utilized in the reverse direction: instead of unrolling a particular iterative algorithm into a network, we can interpret a conventional neural network as a certain iterative algorithm to be identified. Fig.~\ref{fig:reinterp} provides a visual illustration of applying this technique to MLP.\@ Many traditional iterative algorithms have a fixed pattern in 
their iteration steps: a linear mapping followed by a non-linear operation. Therefore, the abstract algorithm in Fig.~\ref{fig:reinterp} represents a broad class of iterative algorithms, which in turn can be identified as deep networks with similar structure to MLPs.\@ The same technique is applicable to other networks, such as CNN or RNN, by replacing the linear operations with convolutions, or by adopting shared parameters across different layers.\@

By interpreting popular network architectures as conventional iterative
algorithms, better understanding of the network behavior and mechanism can be
obtained. Furthermore, rigorous theoretical analysis of designed networks may be facilitated once an equivalence is established with a well-understood class of iterative algorithms.  Finally,
architectural enhancement and performance improvements of the neural networks
may result from incorporating domain knowledge associated with iterative techniques.

\begin{table*}
	\centering
	\caption{Selected Results on Running Time and Parameter Count of Recent Unrolling Works and Alternative Methods.}\label{tab:efficiency}
	\begin{tabularx}{\textwidth}{X X X X}
		\toprule
		& Unrolled Deep Networks & Traditional Iterative Algorithms & Conventional Deep Networks\\
		\midrule
		Reference & Yang {\it et al.}~\cite{yang_admm_csnet} & Metzler {\it et al.}~\cite{metzler_learned_2017} & Kulkarni {\it et al.}~\cite{kulkarni_2016_reconnet}\\
		\midrule
		Running Time (sec) & $2.61$ & $12.59$ & $2.83$\\
		\midrule
		Parameter Count & $7.8\times 10^4$ & $-$ & $3.2\times 10^5$\\
		\bottomrule
		Reference & Solomon {\it et al.}~\cite{solomon_deep_2018} & Beck {\it et al.}~\cite{beck2009fast} & He {\it et al.}~\cite{he_deep_2016}\\
		\midrule
		Running Time (sec) & $5.07$ & $15.33$ & $5.36$\\
		\midrule
		Parameter Count & $1.8\times 10^3$ & $-$ & $8.3\times 10^3$\\
		\bottomrule
		Reference & Li {\it et al.}~\cite{li_2019_deep} & Perrone {\it et al.}~\cite{perrone_clearer_2016} & Kupyn {\it et al.}~\cite{kupyn_deblurgan_2018}\\
		\midrule
		Running Time (sec) & $1.47$ & $1462.90$ & $10.29$\\
		\midrule
		Parameter Count & $2.3\times 10^4$ & $-$ & $1.2\times 10^7$\\
		\bottomrule
	\end{tabularx}
\end{table*}

\begin{figure}
	\includegraphics[width=\linewidth]{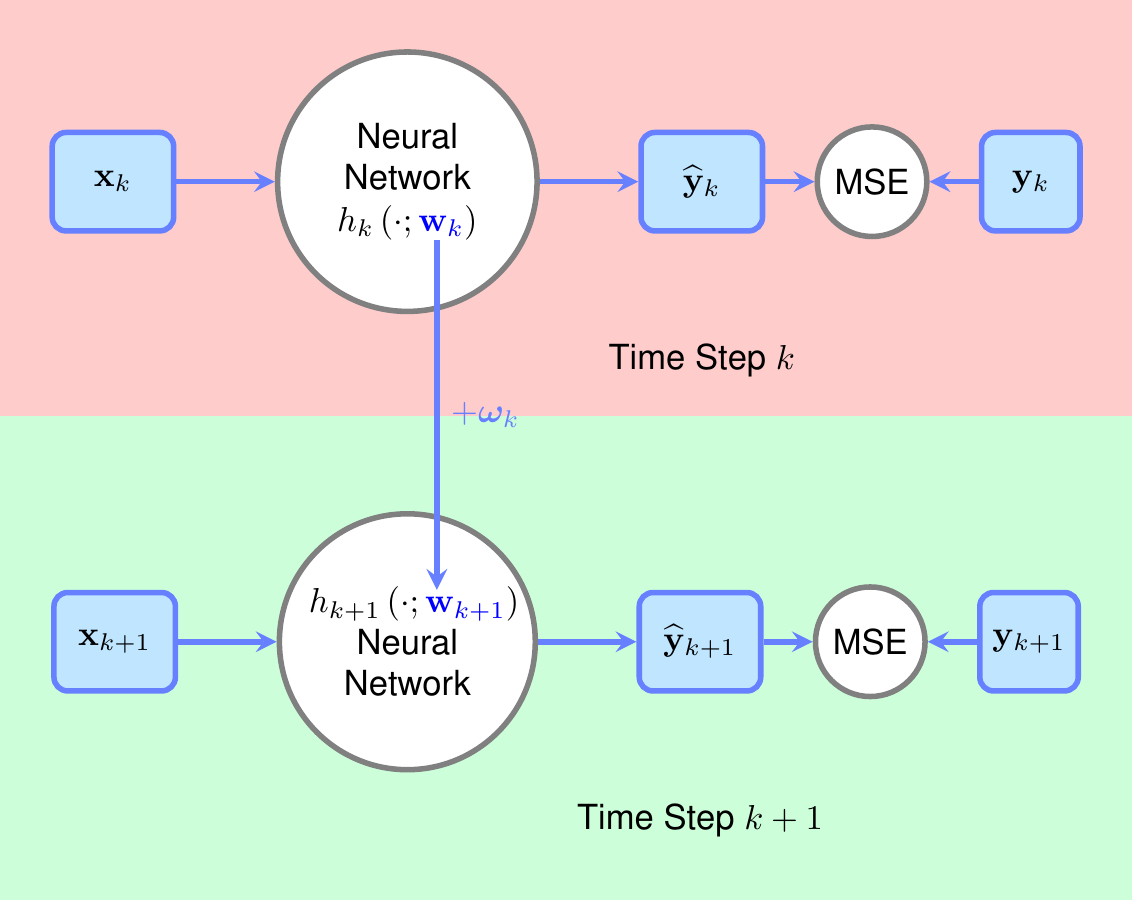}
\caption{Visual illustration of the state-transition model for neural network training. The training data can be viewed as sequentially feeding through the neural network, and the network parameters can be viewed as system states. For more details, refer to the box ``Neural Network Training Using Extended Kalman Filter''.}\label{fig:kalman}
\end{figure}

\begin{strip}
\begin{tcolorbox}[title={Neural Network Training Using Extended Kalman Filter},parbox=false]
	\begin{multicols}{2}
		The Kalman filter is a fundamental technique in signal processing with a wide range of applications. It obtains the Minimum Mean-Square-Error (MMSE) estimation of a system state by recursively drawing observed samples and updating the estimate. The Extended Kalman Filter (EKF) extends to the nonlinear case through iterative linearization. Previous studies~\cite{puskorius_1994_neuro} have revealed that EKF can be employed to facilitate neural network training, by realizing that neural network training is essentially a parameter estimation problem. More specifically, the training samples may be treated as observations, and if the MSE loss is chosen then network training essentially performs MMSE estimation conditional on observations.

		Let $\{(\bx_1,\by_1),(\bx_2,\by_2),\dots,(\bx_N,\by_N)\}$ be a collection of training pairs. We view the training samples as sequentially observed data following a time order. At time step $k$, when feeding $\bx_k$ into the neural network with parameters $\bw$, it performs a nonlinear mapping $h_k(\cdot; \bw_k)$ and outputs an estimate $\widehat{\by}_k$ of $\by_k$. This process can be formally described as the following nonlinear state-transition model:
		\begin{align}
			\bw_{k+1}&=\bw_k+\bm{\omega}_k,\label{eq:process}\\
			\by_k&=h_k\left(\bx_k;\bw_k\right)+\bm{\nu}_k,\label{eq:measurement}
		\end{align}
		where $\bm{\omega}_k$ and $\bm{\nu}_k$ are zero-mean white Gaussian noises with covariance $\cE(\bm{\omega}_k\bm{\omega}_l^T)=\delta_{k,l}\bQ_k$ and $\cE(\bm{\nu}_k\bm{\nu}_l^T)=\delta_{k,l}\bR_k$, respectively. Here $\cE$ is the expectation operator and $\delta$ is the Kronecker delta function. In~\eqref{eq:process}, the noise $\bm{\omega}$ is added artificially to avoid numerical divergence and poor local minima~\cite{puskorius_1994_neuro}. For a visual depiction, refer to Fig.~\ref{fig:kalman}.

		The state-transition model~\eqref{eq:process} and~\eqref{eq:measurement} is a special case of the state-space model of EKF and thus we can apply the EKF technique to estimate the network parameters $\bw_k$ sequentially. To begin with, at $k=0$, $\widehat{\bw}_0$ and $\bP_0$ are initialized to certain values. At time step $k$ ($k\geq 0$), the nonlinear function $h_k$ is linearized as:
		\begin{equation}
			h_k(\bx_k;\bw_k)\approx h_k(\bx_k;\widehat{\bw_k}) + \bH_k(\bw_k-\widehat{\bw}_k),\label{eq:ekf_linear}
		\end{equation}
		where $\bH_k=\left.\frac{\partial h_k}{\partial\bw_k}\right|_{\bw_k=\widehat{\bw}_k}$. For a neural network, $\bH_k$ is essentially the derivative of its output $\widehat{\by}_k$ over its parameters $\bw_k$ and therefore can be computed via back-propagation. The following recursion is then executed:
		\begin{align}
			\bK_k&=\bP_k\bH_k{\left(\bH_k^T\bP_k\bH_k+\bR_k\right)}^{-1},\nonumber\\
			\widehat{\bw}_{k+1}&=\widehat{\bw}_k + \bK_k\left(\by_k-\widehat{\by}_k\right),\label{eq:ekf_update}\\
			\bP_{k+1}&=\bP_k-\bK_k\bH_k^T\bP_k+\bQ_k,\nonumber
		\end{align}
		where $\bK_k$ is commonly called the \emph{Kalman gain}. For details on deriving the update rules~\eqref{eq:ekf_update}, see~\cite[Chapter~1]{haykin_kalman_2001}.

		In summary, neural networks can be trained with EKF by the following steps:
		\begin{enumerate}
			\item Initialize $\widehat{\bw}_0$ and $\bP_0$;
			\item For $k=0, 1, \dots, $
				\begin{enumerate}
					\item Feed $\bx_k$ into the network to obtain the output $\widehat{\by}_k$;
					\item Use back-propagation to compute $\bH_k$ in~\eqref{eq:ekf_linear};
					\item Apply the recursion in~\eqref{eq:ekf_update}.
				\end{enumerate}
		\end{enumerate}

		The matrix $\bP_k$ is the \emph{approximate error covariance matrix}, which models the correlations between network parameters and thus delivers second-order derivative information, effectively accelerating the training speed. For example, in~\cite{singhal_1989_training} it was shown that training an MLP using EKF requires orders of magnitude lower number of epochs than standard back-propagation. 

		In~\cite{haykin_kalman_2001}, some variants of the EKF training paradigm are discussed. The neural network represented by $h_k$ can be a recurrent network, and trained in a similar fashion; the noise covariance matrix $\bR_k$ is scaled to play a role similar to the learning rate adjustment; to reduce computational complexity, a decoupling scheme is employed which divides parameters $\bw_k$ into mutually exclusive groups and turns $\bP_k$ into a block-diagonal matrix.
	\end{multicols}
\end{tcolorbox}
\end{strip}

In this section we will explore close connections between neural networks
and typical families of signal processing algorithms, that are clearly revealed
by unrolling techniques. Specifically, we review studies that reveal the connections between algorithm unrolling and sparse coding, Kalman filters, differential equations and statistical inference from Section~\ref{ssec:sparse} to Section~\ref{ssec:inference} - in that order. We also review selected theoretical advances that
provide formal analysis and rigorous guarantees for unrolling approaches in Section~\ref{ssec:theory}.

\subsection{Connections to Sparse Coding}\label{ssec:sparse}
The earliest work in establishing the connections between neural networks and
sparse coding algorithms dates back to Gregor {\it et
al.}~\cite{gregor_learning_2010}, which we reviewed comprehensively
in Section~\ref{ssec:lista}. A closely related work in the dictionary learning
literature is the task driven dictionary learning algorithm proposed by
Julien {\it et al.}~\cite{mairal_task_2012}. The idea is similar to unrolling:
they view a sparse coding algorithm as a trainable system, whose parameters are
the dictionary coefficients. This viewpoint is equivalent to unrolling the sparse
coding algorithm into a ``network'' of infinite layers, whose output is a limit
point of the sparse coding algorithm. The whole system is trained end-to-end
(task driven) using gradient descent, and an analytical formula for the
gradient is derived.

Sprechmann {\it et al.}~\cite{sprechmann_learning_2015} propose a framework for
training parsimonious models, which summarizes several interesting cases
through an encoder-decoder network architecture. For example, a sparse coding
algorithm, such as ISTA, can be viewed as an encoder as it maps the input
signal into its sparse code. After obtaining the sparse code, the original
signal is recovered using the sparse code and the dictionary. This procedure
can be viewed as a decoder. By concatenating them together, a network is formed
that enables unsupervised learning.  They further extend the model to
supervised and discriminative learning.

Dong {\it et al.}~\cite{dong_super_2016} observe that the forward pass of CNN
basically executes the same operations of sparse-coding based image super
resolution~\cite{yang_scsr_2010}. Specifically, the convolution
operation performs patch extraction, and the ReLU operation mimics sparse
coding. Nonlinear code mapping is performed in the intermediate layers.
Finally, reconstruction is obtained via the final convolution layer. To a certain
extent, this connection explains why CNN has tremendous success in single image
super-resolution.

Jin {\it et al.}~\cite{jin_deep_2017} observe the architectural similarity
between the popular U-net~\cite{Ronneberger_unet_2015} and the unfolded ISTA
network. As sparse coding techniques have demonstrated great success in many
image reconstruction applications such as CT reconstruction, this connection
helps explain why U-net is a powerful tool in these domains, although it was
originally motivated under the context of semantic image segmentation.

\subsection{Connections to Kalman Filtering}\label{ssec:kalman_filter}

Another line of research focuses on acceleration of neural network training by
identifying its relationship with Extended Kalman Filter (EKF). Singhal and
Wu~\cite{singhal_1989_training} demonstrated that neural network training can
be regarded as a nonlinear dynamic system that may be solved by EKF.\@
Simulation studies show that EKF converges much more rapidly than standard
back-propagation. Puskorius and Feldkamp~\cite{puskorius_1994_neuro} propose a
decoupled version of EKF for speed-up, and apply this technique to the training
of recurrent neural networks. More details on establishing the connection
between neural network training and EKF are in the box ``Neural Network
Training Using Extended Kalman Filter''. For a comprehensive review of
techniques employing Kalman filters for network training, refer
to~\cite{haykin_kalman_2001}.

\subsection{Connections to Differential Equations and Variational Methods}\label{ssec:differential}
Differential equations and variational methods are widely applied in numerous
signal and image processing problems. Many practical systems of differential
equations require numerical methods for their solution, and various iterative
algorithms have been developed. Theories around these techniques are extensive
and well-grounded, and hence it is interesting to explore the connections
between these techniques and modern deep learning methods.

In~\cite{chen_trainable_2017}, Chen and Pock adopt the unrolling approach to
improve the performance of Perone-Malik anisotropic
diffusion~\cite{perona_scale-space_1990}, a well-known technique for image
restoration and edge detection. After generalizing the nonlinear diffusion
model to handle non-differentiability, they unroll the iterative discrete
partial differential equation solver, and optimize the filter coefficients and
regularization parameters through training. The trained system proves to be
highly effective in various image reconstruction applications, such as image
denoising, single image super-resolution, and JPEG deblocking.

Recently, Chen {\it et al.\/}~\cite{chen2018neural} identify the residual
layers inside the well-known ResNet~\cite{he_deep_2016} as one iteration of
solving a discrete Ordinary Differential Equation (ODE), by employing the
explicit Euler method. As the time step decreases and the number of layers
increases, the neural network output approximates the solution of the initial
value problem represented by the ODE.\@ Based on this finding they replace the
residual layer with an ODE solver, and analytically derive associated
back-propagation rules that enable supervised learning. In this way, they
construct a network of ``continuous'' depth, and achieve higher parameter
efficiency over conventional ResNet.

The same idea can be applied to other deep learning techniques such as
normalizing flows~\cite{rezende_variational_2015}. Normalizing flows is a
framework for generative modeling, which essentially performs transformations
of random variables belonging to relatively simple distributions to model
complex probability distributions. Typically, the random variables are indexed
by discrete time steps, corresponding to a discrete set of network layers. By
applying the continuation technique, the variable transformation becomes
continuous-in-time, and computation of the expensive log-determinant becomes
unnecessary, leading to significant computational savings.

In physics, Partial Differential Equations (PDE) are frequently used to capture
the dynamics of complex systems. By recasting a generic PDE into a trainable
network, we can discover the underlying physical laws through training. Long
{\it et al.}~\cite{long_pde-net_2018} adopt this principle by approximating
differential operators as convolutional kernels and the nonlinear response
function as a point-wise neural network. In doing so, the model inherits the
predictive power of deep learning systems and the transparency of numerical
PDEs. As a recent follow-up, Long {\it et al.}~\cite{long_pde-net_2019} impose
more constraints on the learnable filters and introduce a symbolic neural
network to approximate the unknown response function.

\subsection{Connections to Statistical Inference and Sampling}\label{ssec:inference}
Statistical inference is broadly defined as the process of drawing conclusions
about populations or scientific truths from data. Popular statistical inference
techniques, such as linear and graphical models, Bayesian inference and Support
Vector Machines (SVM), have demonstrated tremendous successes and effectiveness
in a variety of practical domains. High-impact signal processing and machine
learning applications that involve statistical inference include signal
classification, image reconstruction, representation learning, and more.

Markov Random Field (MRF), as one of the most important graphical models, has
been broadly applied in various image reconstruction and labeling tasks. In its
underlying graph representation, pixels in an image are generally considered
graph nodes whereas their interactions are captured by graph edges.
Traditionally, for tractability only local interactions between spatially
neighboring pixels are modelled, at the sacrifice of model generality and
representation accuracy. Sun and Tappen~\cite{sun_learning_2011} propose to
involve non-local neighbors by grouping similar patches. The generalized MRF
model is called Non-Local Range MRF (NLR-MRF). Inference of NLR-MRF can be
carried out by $K$-steps of gradient descent procedures over the energy
function, which can be unrolled into a $K$-layer deep network. The output of
the network, as a function of the model parameters, is then plugged into the
(empirical) loss function. In this way, end-to-end training can be performed to
optimize the model parameters. In~\cite{sun_learning_2011} this technique
proves its effectiveness in image inpainting and denoising. Specifically,
NLR-MRF demonstrates clear improvements over methods which merely capture local
interactions, and shows on-par performance with state-of-the-art methods.

In a similar spirit, Stoyanov {\it et al.}~\cite{stoyanov_empirical_2011} and
Domke~\cite{domke_parameter_2011} adopt the message passing algorithm, a
dedicated algorithm for inference on graphical models, in the inference stage,
as opposed to gradient descent. Truncating the message passing algorithm by
only executing finite iterations can be regarded as performing approximate
inference, and has the potential benefits of computational savings. From a
conceptual perspective, Domke~\cite{domke_generic_2012} considers abstract
optimization techniques for energy minimization, and focuses on the scenario
where the optimization algorithm runs a fixed number of steps. Compared with
the traditional implicit differentiation approach, this scheme has
computational advantage for large-scale problems because the Hessian matrices
need not be computed. For concrete examples, Domke studies and compares
gradient descent, heavy-ball and Limited-memory
Broyden–Fletcher–Goldfarb–Shanno (LBFGS) algorithms for image labeling and
denoising applications.

Expectation-Maximization (EM) is one of the best known and widely used
techniques in statistical inference. An EM algorithm is an iterative method to
find maximum likelihood or maximum a posteriori (MAP) estimates of parameters
in statistical models, particularly mixture models. For unsupervised perceptual
grouping of image objects, Greff {\it et al.}~\cite{greff_neural_2017} models
the image as a parametrized spatial mixture of $K$ components. They plug-in a
neural network as a transformer which maps the mixture parameters to
probability distributions and hence allows for spatially varying conditional
distributions of image pixels. By employing the EM algorithm and unrolling it
into a recurrent network, an end-to-end differentiable clustering procedure
called N-EM is obtained. A dedicated training technique, called RNN-EM, is also
developed. After training, N-EM is capable of learning how to group pixels
according to constituent objects, promoting localized representation for
individual entities.

In the generative model setting, when the underlying data distribution is
supported on low-dimensional manifolds (a common phenomenon for natural
signals), it has been recognized that entropic metrics induced by maximum
likelihood estimation are fundamentally flawed in principle and perform poorly
in practice. To overcome this issue, metrics based on Optimal Transport (OT)
have become popular choices when measuring the distances between probability
distributions. As an early example, the Wasserstein distance is used as the
loss function in Generative Adversarial Network (GAN) in the seminal work of
Arjovsky {\it et al.}~\cite{arjovsky_wgan_2017}. However, such metrics are
typically defined in variational forms instead of closed forms, and calculating
their derivatives can be problematic. To a large extent this limitation hinders
applications of gradient-based learning techniques.

Recently, algorithm unrolling has become a crucial technique for efficient
minimization of OT-based metrics. In particular, Genevay {\it et
al.}~\cite{genevay_learning_2018} discretize the OT-based loss by drawing
samples, and regularize it with an entropy penalty. The approximate loss is
typically called Sinkhorn loss, and can be computed by the Sinkhorn algorithm.
Genevay {\it et al.} further approximate it by iterating $L$ steps only and
unrolling the Sinkhorn algorithm into $L$ network layers. As each iteration of
the Sinkhorm algorithm is differentiable, the entire network can be trained
end-to-end. In a similar spirit, Patrini {\it et
al.}~\cite{patrini_sinkhorn_2019} employ Wasserstein distance on the latent
space of an autoencoder, and approximate it by $L$ layers of Sinkhorn
iterations. The autoencoder, in combination with these layers, is called
Sinkhorn AutoEncoder (SAE). Patrini {\it et al.} further corroborate the
approximation scheme through theoretical analysis, and experimentally verify
the superior efficiency of Sinkhorn algorithm over the exact Hungarian
algorithm. In unsupervised representation learning experiments, SAE generates
samples of higher quality than other variants of autoencoders, such as
variational autoencoder~\cite{kingma_auto-encoding_2014} and Wasserstein
autoencoder~\cite{tolstikhin_wae_2018}.

\subsection{Selected Theoretical Studies}\label{ssec:theory}
Although LISTA successfully achieves higher efficiency than the iterative
counterparts, it does not necessarily recover a more accurate sparse code
compared to the iterative algorithms, and thorough theoretical analysis of its
convergence behavior is yet to be developed.

Xin {\it et al.}~\cite{xin2016maximal} study the unrolled Iterative Hard
Thresholding (IHT) algorithm, which has been widely applied in $\ell^0$ norm
constrained estimation problems and resembles ISTA to a large extent. The
unrolled network is capable of recovering sparse signals from dictionaries with
coherent columns. Furthermore, they analyze the optimality criteria for the
network to recover the sparse code, and verify that the network can achieve
linear convergence rate under appropriate training.

In a similar fashion, Chen {\it et al.}~\cite{chen_theoretical_2018} establish
a linear convergence guarantee for the unrolled ISTA network. They also derive
a weight coupling scheme similar to~\cite{xin2016maximal}. As a follow-up, Liu
{\it et al.}~\cite{liu2018alista} characterize optimal network parameters
analytically by imposing mutual incoherence conditions on the network weights.
Analytical derivation of the optimal parameters help reduce the parameter
dimensionality to a large extent. Furthermore, they demonstrate that a network
with analytic parameters can be as effective as the network trained completely
from data. For more details on the theoretical studies around LISTA,
refer to the box ``Convergence and Optimality Analysis of LISTA''.

Papyan {\it et al.}~\cite{papyan_convolutional_2017} interpret CNNs as executing finite iterations of the
Multi-Layer Convolutional Sparse Coding (MLCSC) algorithm. In other words, CNN
can be viewed as an unrolled ML-CSC algorithm. In this interpretation, the
convolution operations naturally emerge out of a convolutional sparse
representation, with the commonly used soft-thresholding operation viewed as a
symmetrized ReLU.\@ They also analyze the ML-CSC problem
and offer theoretical guarantees such as uniqueness of the multi-layer sparse
representation, stability of the solutions under small perturbations, and
effectiveness of ML-CSC in terms of sparse recovery. In a recent follow-up work~\cite{sulam_2019_mlfista}, they further propose dedicated iterative optimization algorithms for
solving the ML-CSC problem, and demonstrate superior efficiency over other conventional algorithms such as ADMM and FISTA for solving the multi-layer bais purisuit problem.

\section{Perspectives and Recent Trends}\label{sec:discussions}
We reflect on the remarkable effectiveness of algorithm unrolling in Section~\ref{ssec:general_pespective}. Recent trends and current concerns regarding algorithm unrolling are discussed in Section~\ref{ssec:trends}. We contrast algorithm unrolling with alternatives and discuss their relative merits and drawbacks in Section~\ref{subsec:alternatives}.

\subsection{Distilling the Power of Algorithm Unrolling}\label{ssec:general_pespective}
In recent years, algorithm unrolling has proved highly effective in achieving
superior performance and higher efficiency in many practical domains. A
question that naturally arises is, why is it so powerful?

\begin{figure}
	\centering
	\includegraphics[width=\linewidth]{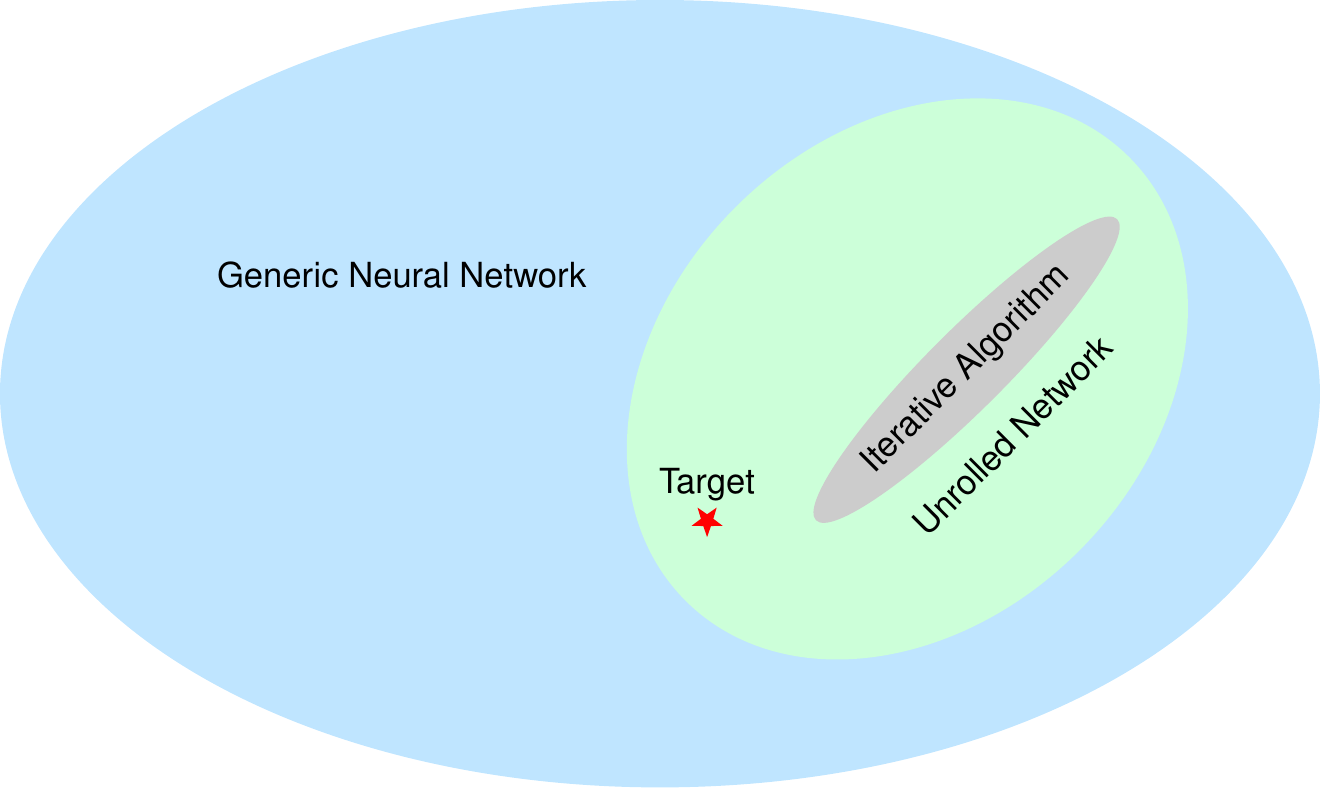}
	\caption{A high-level unified interpretation of algorithm unrolling from a functional approximation perspective: the ellipses depict the scope of functions that can be approximated by each category of methods. Compared with iterative algorithms which have limited representation power and usually underfit the target, unrolled networks usually better approximate the target thanks to its higher representation power. On the other hand, unrolled networks have lesser representation power than generic neural networks but usually generalize better in practice, hence providing an attractive balance.}\label{fig:unify}
\end{figure}

Fig.~\ref{fig:unify} provides a high-level illustration of how algorithm
unrolling can be advantageous compared with both traditional iterative
algorithms and generic neural networks, from a functional approximation
perspective. By parameter tuning and customizations, a traditional iterative
algorithm spans a relatively small subset of the functions of interest, and
thus has limited representation power. Consequently, it is capable of
approximating a given target function~\footnote{We use the phrase ``target function'' to refer to typical functional mappings that model the relationship between a given input and desired output in real-world problems such as signal recovery and classification.} reasonably well, while still leaving some
gaps that undermine performance in practice. Nevertheless, iterative algorithms generalize relatively well in limited training scenarios. From a statistical learning
perspective, iterative algorithms correspond to models of high bias but low
variance.

On the other hand, a generic neural network is capable of more accurately
approximating the target function thanks to its universal approximation
capability. Nevertheless, as it typically consists of an enormous number of
parameters, it constitutes a large subset in the function space. Therefore, when
performing network training the search space becomes large and training is
a major challenge. The high dimensionality of parameters also requires an abundant of training samples and generalization becomes
an issue. Furthermore, network efficiency may also suffer as the network size
increases. Generic neural networks are essentially models of high variance but
low bias.

In contrast, the unrolled network, by expanding the capacity of iterative
algorithms, can approximate the target function more accurately, while spanning
a relatively small subset in the function space. Reduced size of the search
space alleviates the burden of training and requirement of large scale training
datasets. Since iterative algorithms are carefully developed based
on domain knowledge and already provide reasonably accurate approximation of
the target function, by extending them and training from real data unrolled
networks can often obtain highly accurate approximation of the target
functions. As an intermediate state between generic networks and iterative
algorithms, unrolled networks typically have relatively low bias and variance
simultaneously. Table~\ref{tab:feature_comp} summarizes some features of
iterative algorithms, generic networks and unrolled networks.

\begin{strip}
	\begin{tcolorbox}[title={Convergence and Optimality Analysis of LISTA},parbox=false]
		\begin{multicols}{2}
			Although it is shown in~\cite{gregor_learning_2010} that LISTA
			achieves empirically higher efficiency than ISTA through training,
			several conceptual issues remain to be addressed. First, LISTA does
			not exhibit superior performance over ISTA, not even under
			particular scenarios; second, the convergence rate of LISTA is
			unknown; third, LISTA actually differs from ISTA by introducing
			artificial parameter substitutions; and finally, the optimal
			parameters are learned from data, and it is difficult to have a
			sense of what they look like.

			To address these open issues, several recent theoretical studies
			have been conducted. A common assumption is that there exists a
			sparse code $\bx^\ast\in\mathbb{R}^m$ which approximately satisfies
			the linear model $\by\approx\bW\bx^\ast$ where
			$\bW\in\mathbb{R}^{n\times m}$ and $m>n$. More specifically, it is
			commonly assumed that $\|\bx^\ast\|_0\leq s$ for some positive
			integer $s$, where $\|\cdot\|_0$ counts the number of nonzero
			entries.

			Xin {\it et al.}~\cite{xin2016maximal} examine a closely related sparse coding problem:
			\begin{equation}
				\min_{\bx}\frac{1}{2}\|\by-\bW\bx\|_2^2\quad\text{subject to}\quad\|\bx\|_0\leq k,\label{eq:l0_min}
			\end{equation}
			where $k$ is a pre-determined integer to control the sparsity level of $\bx$. In~\cite{xin2016maximal} a network is constructed by unrolling the Iterative Hard-Thresholding (IHT) algorithm, which has similar form as ISTA.\@ At layer $l$, the following iteration is performed:
			\begin{equation}
				\bx^{l+1}=\cH_k\left\{\bW_t\bx^l+\bW_e\by\right\},
			\end{equation}
			where $\cH_k$ is the hard-thresholding operator which keeps $k$ coefficients of largest magnitude and zeroes out the rest. Xin {\it et al.}~\cite{xin2016maximal} proved that in order for the IHT-based network to recover $\bx^\ast$, it must be the case that 
			\begin{equation}
				\bW_t=\bI-\bm{\Gamma}\bW,
			\end{equation}
			for some matrix $\bm{\Gamma}$, which implies that the implicit variable substitution $\bW_t=\bI-\frac{1}{\mu}\bW^T\bW$ and $\bW_e=\frac{1}{\mu}\bW^T$ may not play such a big role as it seems, as long as the network is trained properly so that it acts as a generic sparse recovery algorithm. Furthermore, Xin {\it et al.} showed that upon some modifications such as using layer-specific parameters, the learned network can recover the sparse code even when $\bW$ admits correlated columns, a scenario known to be particularly challenging for traditional iterative sparse coding algorithms.

			Chen {\it et al.}~\cite{chen_theoretical_2018} perform similar analysis on LISTA with layer-specific parameters, i.e., in layer $l$ the parameters $\left(\bW_t^l,\bW_e^l,\lambda^l\right)$ are used instead. Similar to Xin {\it et al.}~\cite{xin2016maximal}, they also proved that under certain mild assumptions, whenever LISTA recovers $\bx^\ast$, the following weight coupling scheme must be satisfied asymptotically:
			\begin{equation*}
				\bW_t^l-\left(\bI-\bW_e^l\bW\right)\rightarrow 0,\quad\text{as }l\rightarrow\infty,
			\end{equation*}
			which shows that the implicit variable substitutions may be inconsequential in an asymptotic sense. Therefore, they adopted the following coupled parameterization scheme:
			\begin{equation*}
				\bW_t^l=\bI-\bW_e^l\bW,
			\end{equation*}
			and proved that the resulting network recovers $\bx^\ast$ in a linear rate, if the parameters ${\left(\bW_e^k,\lambda_k\right)}_{k=1}^\infty$ are appropriately selected. They further integrate a support selection scheme into the network. The network thus has both weight coupling and support selection structures and is called LISTA-CPSS.\@

			Liu {\it et al.}~\cite{liu2018alista} extend Chen {\it et
			al.}~\cite{chen_theoretical_2018}'s work by analytically
			characterizing the optimal weights $\bW_e^k$ for LISTA-CPSS.\@ They
			proved that, under certain regularity conditions, a linear
			convergence rate can be achieved if ${\left(\bW_e^k,
			\lambda^k\right)}_k$ are chosen in a specific form. This implies
			that the network with analytic parameters can be asymptotically as efficient as
			the trained version. Although the analytic forms may be nontrivial
			to compute in practice, their analysis helps reducing the number of
			network parameters dramatically.
		\end{multicols}
	\end{tcolorbox}
\end{strip}

\begin{table*}
	\centering
	\caption{Feature Comparisons of Iterative Algorithms, Generic Deep Networks, and Unrolled Networks.}\label{tab:feature_comp}
	\begin{tabularx}{\textwidth}{X c X X X X}
		\toprule
		Techniques & Performance & Efficiency & Parameter Dimensionality & Interpretability & Generalizability\\
		\midrule
		Iterative Algorithms & Low & Low & Low & High & High \\
		\midrule
		Generic Deep Networks  & High & High & High & Low & Low\\
		\midrule
		Unrolled Networks & High & High & Middle & High & Middle\\
		\bottomrule
	\end{tabularx}
\end{table*}

\subsection{Trends: Expanding Application Landscape and Addressing Implementation Concerns}\label{ssec:trends}
A continuous trend in recent years is to explore more general underlying
iterative algorithms. Earlier unrolling approaches were centered around the ISTA
algorithm~\cite{gregor_learning_2010,jin_deep_2017,wang_deep_2015}, while
recently other alternatives have been pursued such as Proximal
Splitting~\cite{sprechmann_learning_2015}, ADMM~\cite{yang_admm_csnet}, Half
Quadratic Splitting~\cite{li_icassp19}, to name a few. For instance, Metz {\it et al.}~\cite{metz_unrolled_2016} unroll the ADAM optimizer~\cite{kingma_adam:_2015} to stabilize the GAN training, while Diamond {\it et al.}~\cite{diamond_unrolled_2018} propose a general framework for unrolled optimization. Consequently, a growing
number of unrolling approaches as well as novel unrolled network architectures
appear in recent publications.

In addition to expanding the methodology, researchers are broadening the
application scenarios of algorithm unrolling. For instance, in communications
Samuel {\it et al.}~\cite{samuel_2017_deep} propose a deep network called
DetNet, based on unrolling the projected gradient descent algorithm for least
squares recovery. In multiple-input-multiple-output detection tasks, Det-Net
achieves similar performance to a detector based on semidefinite relaxation,
while being than 30 times faster. Further, DetNet exhibits promising
performance in handling ill-conditioned channels, and is more robust than the
approximate message passing based detector as it does not require knowledge of
the noise variance. More examples of unrolling techniques in
communications can be found in~\cite{balatsoukas2019deep, farsad_data-driven_2020}.

From an optimal control viewpoint, Li {\it et al.}~\cite{li2017maximum}
interpret deep neural networks as dynamic systems and recast the network
training procedure as solving an optimal control problem. By analyzing the
corresponding Pontryagin's maximum principle, they devise a novel network
training algorithm. Compared with conventional gradient-based methods, the
proposed algorithm has faster initial convergence and is resilient to stalling
in flat landscape. The principles of optimal control  have also inspired
researchers towards the design of real world imaging systems. In such an
approach to image restoration, Zhang {\it et al.}~\cite{zhang_dynamically_2018}
argue that different end points must be chosen when handling images of
different degradation levels. To this end, they introduce a dedicated policy
network for predicting the end point. The policy network is essentially a
convolutional RNN.\@ The estimated end point is used to govern the termination
of the restoration network. Both networks interplay and are trained under a
reinforcement learning framework. The entire model is thus called Dynamically
Unfolding Recurrent Restorer (DURR). Experiments on blind image denoising and
JPEG deblocking verify that DURR is capable of delivering higher quality
reconstructed images with sharper details, and generalizes better when the
degradation levels vary or are unseen in the training datasets compared with
its competitors. Furthermore, DURR has significantly fewer parameters and
higher runtime efficiency.

The unrolled network can share parameters across all the layers, or carry over
layer-specific parameters. In the former case, the networks are typically more
parameter efficient. However, how to train the network effectively is a
challenge because the networks essentially resemble RNNs and may similarly
suffer from gradient explosion and vanishing problems. In the latter case, the networks
slightly deviate from the original iterative algorithm and may not completely
inherit its theoretical benefits such as convergence guarantees. However, the
networks can have enhanced representation power and adapt to real world
scenarios more accurately. The training may also be much easier compared with
RNNs. In recent years, a growing number of unrolling techniques allow the
parameters to vary from layer to layer.

An interesting concern relates to deployment of neural networks on
resource-constrained platforms, such as digital single-lens reflex cameras and mobile devices. The heavy
storage demand renders many top-performing deep networks impractical, while
straightforward network compression usually deteriorates its performance
significantly. Therefore, in addition to computational efficiency, nowadays
researchers are paying increasing attention to the parameter efficiency aspect,
and increasing research attention is paid to algorithm unrolling.

Finally, there are other factors to be considered when constructing unrolled
networks.  In particular, many iterative algorithms when unrolled
straightforwardly may introduce highly non-linear and/or non-smooth operations
such as hard-thresholding. Therefore, it is usually desirable to design
algorithms whose iteration procedures are either smooth or can be well
approximated by smooth operators. Another aspect relates to the network depth.
Although deeper networks offer higher representation power, they are
generally harder to train in practice~\cite{he_deep_2016}. Indeed, techniques
such as stacked pre-training have been frequently employed in existing
algorithm unrolling approaches to overcome the training difficulty to some
extent. Taking this into account, iterative algorithms with faster convergence
rate and simpler iterative procedures are generally considered more often.

\subsection{Alternative Approaches}\label{subsec:alternatives}
\begin{figure}
	\includegraphics[width=\linewidth]{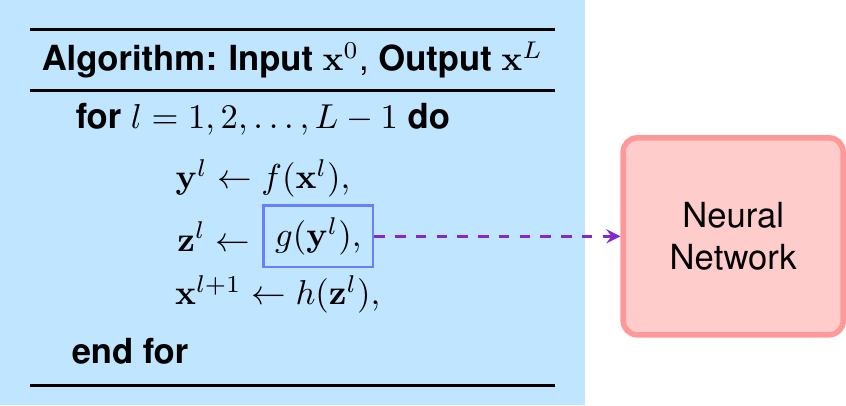}
	\caption{An alternative approach to algorithm unrolling is to replace one step of the iterative algorithm with an intact conventional neural network.}\label{fig:plugin_in}
\end{figure}

Besides algorithm unrolling, there are other approaches for characterizing or
enhancing interpretability of deep networks. The initial motivation of neural
networks is to model the behavior of the human brain. Traditionally neural networks are
often interpreted from a neurobiological perspective. However, discrepancies
between actual human brain and artificial neural networks have been constantly
observed. In recent years, there are other interesting works on identifying and
quantifying network interpretability by analyzing the correlations between
neuron activations and human perception. One such example is the emerging
technique called ``network dissection''~\cite{zhou_interpret_2019} which
studies how the neurons capture semantic objects in the scene and how
state-of-the art networks internally represent high-level visual concepts.
Specifically, Zhou {\it et al.}~\cite{zhou_interpret_2019} analyze the neuron activations on
pixel-level annotated datasets, and quantify the network interpretability by
correlating the neuron activations with groundtruth annotations. Bau {\it et
al.}~\cite{bau2019gandissect} extends this technique to generative adversarial
networks. These works complement algorithm unrolling by offering
visual and biological interpretations of deep networks. However, they are
mainly focused on characterizing the interpretability of existing networks and
are less effective at connecting neural networks with traditional iterative
algorithms and motivating novel network architectures.

Another closely related technique is to employ a conventional deep network as
drop-in replacement of certain procedures in an iterative algorithm.
Fig.~\ref{fig:plugin_in} provides a visual illustration of this technique. The
universal approximation theorem~\cite{cybenko_approximation_1989} justifies the
use of neural networks to approximate the algorithmic procedures, as long as
they can be represented as continuous mappings. For instance,
in~\cite{metzler_learned_2017} Metzler {\it et al.\/} observe that one step of
ISTA may be treated as a denoising procedure, and henceforth can be replaced by
a denoising CNN\@. The same approach applies to Approximate Message Passing
(AMP)~\cite{donoho2009message}, an extension to ISTA\@. Meinhardt {\it et
al.}~\cite{meinhardt_learning_2017} replace the proximal operator of the
regularization  used in many convex energy minimization algorithms with a
denoising neural network. In this way, the neural network acts as an implicit
regularizer in many inverse problems, or equivalently a natural image prior.
The denoising neural network can then be employed in different applications,
alleviating the need for problem-specific training. In a similar fashion,
in~\cite{gupta2018cnn} Gupta {\it et al.\/} replace the projection procedure in
projected gradient descent with a denoising CNN\@. Shlezinger {\it et
al.}~\cite{shlezinger_2019_viterbinet} replace the evaluation of log-likelihood
in the Viterbi algorithm with dedicated machine learning methods, including a
deep neural network.  Ryu {\it et al.}~\cite{ryu_plug-and-play_2019} prove
that, when the denoisers satisfy certain Lipschitz conditions, replacing the
proximal operators with denoisers leads to convergence for some popular
optimization algorithms such as ADMM and forward-backward splitting. Based on
this theoretical result, they also developed a technique to enforce the
Lipschitz conditions when training the denoisers.

This technique has the advantage of inheriting the knowledge about conventional
deep networks, such as network architectures, training algorithms,
initialization schemes, etc. In addition, in practice this technique can
effectively complement the limitations of iterative algorithms. For instance,
Shlezinger {\it et al.}~\cite{shlezinger_2019_viterbinet} demonstrated that, by
replacing part of the Viterbi algorithm with a neural network, full knowledge
about the statistical relationship between channel input and output is no
longer necessary. Therefore, the resulting algorithm achieves higher robustness
and better performance under model imperfections. Nevertheless, the procedures
themselves are still approximated abstractly via conventional neural networks.

\section{Conclusions}\label{ssec:limitations}
In this article we provide an extensive review of algorithm unrolling, starting with LISTA as a basic example. We then showcased practical applications of unrolling in various real-world signal and image processing problems. In many application domains, the unrolled interpretable deep networks offer state-of-the art performance, and achieve high computational efficiency. From a conceptual standpoint, algorithm unrolling also helps reveal the connections between deep learning and other important categories of approaches, that are widely applied for solving signal and image processing problems.

Although algorithm unrolling is a promising technique to build efficient and
interpretable neural networks, and has already achieved success in many
domains, it is still evolving. We conclude this article by discussing limitations and
open challenges related to algorithm unrolling, and suggest possible directions
for future research.

\emph{Proper Training of the Unrolled Networks:}
The unrolling techniques provide a powerful principled framework for
constructing interpretable and efficient deep networks; nevertheless, the full
potential of  unrolled networks can be exploited only when they are trained
appropriately. Compared to popular conventional networks (CNNs,
auto-encoders), unrolled networks usually exhibit customized structures.
Therefore, existing training schemes may not work well. In addition, the
unrolled network sometimes delivers shared parameters among different layers,
and thus it resembles RNN, which is well-known to be difficult to
train~\cite{pascanu_difficulty_2013}. Therefore, many existing works apply
greedy layer-wise pre-training. The development of well-grounded end-to-end training schemes for unrolled networks
continues to be a topic of great interest.

A topic of paramount importance is how to initialize the network. While there
are well studied methods for initializing conventional
networks~\cite{glorot_understanding_2010,he_delving_2015}, how to
systematically transfer such knowledge to customized unrolled networks remains
a challenge. In addition, how to prevent vanishing and exploding gradients
during training is another important issue. Developing equivalents or counterparts of established practices such as
Batch Normalization~\cite{ioffe2015batch} and Residual
Learning~\cite{he_deep_2016} for unrolled networks is a viable research direction.

\emph{Bridging the Gap between Theory and Practice:}
While substantial progress has been achieved towards understanding the network
behavior through unrolling, more works need to be done to thoroughly understand
its mechanism. Although the effectiveness of some networks on image
reconstruction tasks has been explained somehow by drawing parallels to
sparse coding algorithms, it is still mysterious why state-of-the art networks
perform well on various recognition tasks. Furthermore, unfolding itself is not
uniquely defined. For instance, there are multiple ways to choose the
underlying iterative algorithms, to decide what parameters become trainable and
what parameters to fix, and more. A formal study on how these choices affect
convergence and generalizability can provide valuable insights and practical
guidance.

Another interesting direction is to develop a theory that provides guidance for
practical applications. For instance, it is interesting to perform analysis
that guide practical network design choices, such as dimensions of parameters,
network depth, etc. It is particularly interesting to identify factors that have high
impact on network performance.

\emph{Improving the Generalizability:} One of the critical limitations of common deep networks is its lack of
generalizability, i.e., severe performance degradations when operating on
datasets significantly different from training. Compared with neural networks,
iterative algorithms usually generalize better, and it is interesting to
explore how to maintain this property in the unrolled networks. Preliminary investigations have shown improved generalization experimentally of unrolled networks in a few cases~\cite{li_2019_deep} but a formal theoretic understanding remains elusive and is highly desirable. From an impact standpoint, this line of
research may provide newer additions to approaches for
semi-supervised/unsupervised learning, and offer practical benefits when
training data is limited or when working with resource-constrained platforms.

\bibliographystyle{IEEEbib}
\bibliography{proposal_refs,Deconvolution,DeepLearning}

\begin{IEEEbiography}[{\includegraphics[width=1in,height=1.25in,clip,keepaspectratio]{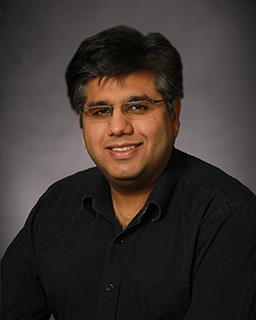}}]{Vishal Monga}
	has been on the EECS faculty at Penn State since Fall 2009. From Oct 2005-July 2009 he was an imaging scientist with Xerox Research Labs. He has also been a visiting researcher at Microsoft Research in Redmond, WA and a visiting faculty at the University of Rochester. Prior to that, he received his PhDEE from the department of Electrical and Computer Engineering at the University of Texas, Austin. Dr.\ Monga is an elected member of the IEEE Image Video and Multidimensional Signal Processing (IVMSP) Technical Committee. He has served on the editorial boards of the IEEE Transactions on Image Processing, IEEE Signal Processing Letters and IEEE Transactions on Circuits and Systems for Video Technology. He is currently a Senior Area Editor for IEEE Signal Processing Letters and the lead Guest Editor for the IEEE Journal of Selected Topics in Signal Processing Special Issue: Domain Enriched Learning for Medical Imaging. He is a recipient of the US National Science Foundation CAREER award and a 2016 Joel and Ruth Spira Teaching Excellence award. His group's research focuses on optimization based methods with applications in signal and image processing, learning and computer vision.
\end{IEEEbiography}

\begin{IEEEbiography}[{\includegraphics[width=1in,height=1.25in,clip,keepaspectratio]{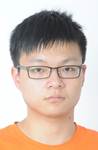}}]{Yuelong Li}
	(S'14) received the B.S. degree in information engineering from Shanghai
	Jiao Tong University, Shanghai, China, in 2013, and the Ph.D. degree in
	electrical engineering from Pennsylvania State University, University Park,
	PA, USA, in 2018. From Nov 2018 to Mar 2020 he was working at Sony US
	Research Center. He is currently an applied scientist at Amazon Lab 126.
	His research interests include computational photography and 3D modeling,
	with a focus on nonlinear programming techniques and more recently deep
	learning techniques.
\end{IEEEbiography}

\begin{IEEEbiography}[{\includegraphics[width=1in,height=1.25in,clip,keepaspectratio]{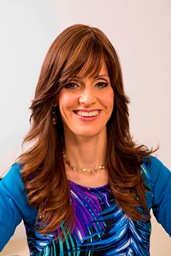}}]{Yonina C. Eldar}
	received the B.Sc.\ degree in Physics in 1995 and the B.Sc.\ degree in Electrical Engineering in 1996 both from Tel-Aviv University (TAU), Tel-Aviv, Israel, and the Ph.D. degree in Electrical Engineering and Computer Science in 2002 from the Massachusetts Institute of Technology (MIT), Cambridge. From January 2002 to July 2002 she was a Postdoctoral Fellow at the Digital Signal Processing Group at MIT.\ She is currently a Professor in the Department of Mathematics and Computer Science, Weizmann Institute of Science, Rehovot, Israel. She was previously a Professor in the Department of Electrical Engineering at the Technion, where she held the Edwards Chair in Engineering. She is also a Visiting Professor at MIT, a Visiting Scientist at the Broad Institute, and an Adjunct Professor at Duke University and was a Visiting Professor at Stanford. She is a member of the Israel Academy of Sciences and Humanities (elected 2017), an IEEE Fellow and a EURASIP Fellow.

She is the Editor in Chief of Foundations and Trends in Signal Processing, a member of the IEEE Sensor Array and Multichannel Technical Committee and serves on several other IEEE committees. In the past, she was a Signal Processing Society Distinguished Lecturer, member of the IEEE Signal Processing Theory and Methods and Bio Imaging Signal Processing technical committees, and served as an associate editor for the IEEE Transactions On Signal Processing, the EURASIP Journal of Signal Processing, the SIAM Journal on Matrix Analysis and Applications, and the SIAM Journal on Imaging Sciences. She was Co-Chair and Technical Co-Chair of several international conferences and workshops. She is author of the book ``Sampling Theory: Beyond Bandlimited Systems'' and co-author of the books ``Compressed Sensing'' and ``Convex Optimization Methods in Signal Processing and Communications'', all published by Cambridge University Press.
\end{IEEEbiography}

\end{document}